\documentclass[12pt,a4paper]{article}
\pdfoutput=1
\usepackage{inputenc}   
\usepackage{amsmath}
\usepackage{amssymb}   
\usepackage{amsbsy}   
\usepackage{graphicx,psfrag,epsf}
\usepackage{enumerate}
\usepackage{multirow}
\usepackage[hang,small,tight]{subfigure}   
\usepackage{indentfirst}   
\usepackage{url} 
\usepackage{hhline}
\usepackage[pdftex,bookmarks=true,colorlinks=false,linkcolor=black,citecolor=black]{hyperref}

\usepackage{appendix}

\newcommand{\bfeta}{\mbox{\boldmath $\eta$}}

\newcommand{\bfrho}{\mbox{\boldmath $\rho$}}

\newcommand{\bfpi}{\mbox{\boldmath $\pi$}}
\newcommand{\bftheta}{\mbox{\boldmath $\theta$}}
\newcommand{\bfTheta}{\mbox{\boldmath $\Theta$}}

\newcommand{\bflambda}{\mbox{\boldmath $\lambda$}}

\newcommand{\bfnu}{\mbox{\boldmath $\nu$}}
\newcommand{\bfomega}{\mbox{\boldmath $\omega$}}

\newcommand{\sbar}{\bar{s}}

\newcommand{\bfc}{\mbox{$\mathbf c$}}
\newcommand{\bfC}{\mbox{$\mathbf C$}}

\newcommand{\bfe}{\mbox{$\mathbf e$}}

\newcommand{\bfG}{\mbox{$\mathbf G$}}

\newcommand{\bfI}{\mbox{$\mathbf I$}}

\newcommand{\bfN}{\mbox{$\mathbf N$}}

\newcommand{\bfv}{\mbox{$\mathbf v$}}
\newcommand{\bfV}{\mbox{$\mathbf V$}}

\newcommand{\bfY}{\mbox{$\mathbf Y$}}

\newcommand{\bfZ}{\mbox{$\mathbf Z$}}

\newcommand{\bfone}{\mbox{$\mathbf 1$}}

\def\spacingset#1{\renewcommand{\baselinestretch}%
	{#1}\small\normalsize} \spacingset{1}

\begin{document}
	
	\title{\bf Model-based Inference for Rare and\\ Clustered Populations from Adaptive\\ Cluster Sampling using Auxiliary Variables}
	\author{Izabel N. de Souza\thanks{Departamento de Estatística, Instituto de Matemática e Estatística, Universidade Federal do Rio de Janeiro (UFRJ), RJ, Brazil, \url{nolau@dme.ufrj.br}} , Kelly C. M. Gonçalves\thanks{Departamento de Estatística, Instituto de Matemática e Estatística, Universidade Federal do Rio de Janeiro (UFRJ), RJ, Brazil, \url{kelly@dme.ufrj.br}} \ and João B. M. Pereira\thanks{Departamento de Estatística, Instituto de Matemática e Estatística, Universidade Federal do Rio de Janeiro (UFRJ), RJ, Brazil, \url{joao@dme.ufrj.br}}}\date{}
	\maketitle
	
	\bigskip
	\begin{abstract}
		Rare populations, such as endangered animals and plants, drug users and individuals with rare diseases, tend to cluster in regions. Adaptive cluster sampling is generally applied to obtain information from clustered and sparse populations since it increases survey effort in areas where the individuals of interest are observed. This work aims to propose a unit-level model which assumes that counts are related to auxiliary variables, improving the sampling process, assigning different weights to the cells, besides referring them spatially. The proposed model fits rare and grouped populations, disposed over a regular grid, in a Bayesian framework. The approach is compared to alternative methods using simulated data and a real experiment in which adaptive samples were drawn from an African Buffaloes population in a 24,108$km^2$ area of East Africa. Simulation studies show that the model is efficient under several settings, validating the methodology proposed in this paper for practical situations. 
	\end{abstract}
	
	\noindent%
	\vfill
	
	\newpage
	\spacingset{1.3}
	\section{Introduction}
	In several statistical surveys, there are obstacles in data collection since the study object is hard to observe either because it is a rare population, exhibits a pattern of sparsely distributed groups in a region, or is mobile over time. Examples of populations with these characteristics include endangered animals and plants, ethnic minorities, drug users, individuals with rare diseases, and recent immigrants. Assume that the population of interest is spatially distributed in a region of interest, where a regular grid with $M$ equal-sized cells is superimposed. Denote the partitioned region by $R=\{c_1,\dots,c_M\}$. Let $\eta(c)$ denote the number of individuals of the population within the grid cell $c$, for all $c \in R$, that is, this cell's count. The objective is to estimate a rare and clustered population total $T=\sum_{c \in R}\eta(c)$.
	
	Under traditional sampling methods of grid cells, a subset of $m<M$ cells is drawn and their respective counts $\eta(c)$ are observed. Due to the population characteristics, small sample sizes result in large numbers of empty grid cells, for which $\eta(c)=0$, leading us to inaccurate estimates of the population quantity of interest. In this context, adaptive cluster sampling, introduced by \cite{Thompson1990}, is a way to surmount this difficulty by increasing survey effort around non-empty grid cells of the sample. From an initial sample of $m$ grid cells, when we find a non-empty grid cell, for which $\eta(c)\neq 0$, we also sample its neighbors and continue surveying until we obtain a set of contiguous non-empty grid cells surrounded by empty grid cells. By this way, empty grid cells bring no further survey effort. Therefore, adaptive cluster sampling requires some prior knowledge about the structure of the subjacent population, which may be obtained from a preliminary survey, to be effective.
	
	According to \cite{Thompson1990}, the set of contiguous non-empty grid cells is called a network; this set plus its neighboring empty grid cells are together named a cluster; and empty cells are defined as one-sized networks. Therefore, $R$ is exhaustively partitioned into disjoint networks, and the final sample contains empty and non-empty networks. \cite{Thompson1990} treated empty edge cells as unobserved and, from an initial random sample without replacement of grid cells, inclusion probabilities are assigned to the sampled networks, used to construct design-unbiased estimators of $T$ and their variances. Note that, the networks are the basis of the analysis and, although the initial selection of cells is without replacement, the same network can be selected more than once, a problem that \cite{Thompson1990} solved by allowing multiple inclusions of networks. Edge cells can be incorporated into the estimator by taking its conditional expectation given the minimal sufficient statistic and setting the Rao-Blackwell improved version of that. These estimators were described and computed for small sample sizes in \cite{Thompson1990}. Further, \cite{Salehi1997} proposed a scheme whereby the networks are selected one by one without replacement, avoiding select the same network more than once.
	
	Several studies have been conducting using adaptive sampling designs on real populations. For example, \cite{Smith1995} studied the methodology for rare species of waterfowl, \cite{Su2003} discussed adaptive cluster sampling with order statistics and a stopping rule for a fish population, \cite{Philippi2005} showed that it is a viable alternative for the estimation of occurences in local populations of low-abundance plants and \cite{Gattone2016} applied it to negative correlated data.
	
	\cite{Thompson1996} examined some general ideas about model-based inference approaches for adaptive sampling. The likelihood-based methods, such as Bayesian estimation, showed promising results among model-based approaches. Beside them, Bayesian inference methods for adaptive cluster sampling designs have been developed in \cite{Rapley2008} and \cite{Goncalves2016}, which incorporate prior knowledge that the population is rare and grouped for both inference and sample design. \cite{Rapley2008} provided a model at the network level, while \cite{Goncalves2016} modeled at the cell level, considering heterogeneity among units belonging to different clusters. Both works did not take into account the spatial locations of the networks, a fact that does not cause any loss of information about the total population since it does not depend on where the networks are located, under the model.
	
	A possible approach to spatially model clustered data is by using point processes (\cite{Diggle1975}; \cite{Baddeley2000}; \cite{Brix2001}), where the clusters are considered as points and have no internal spatial structure, although there is a spatial relationship between them. \cite{Rapley2008} place the clusters and give them a spatial size by superimposing a grid on a region containing a clustered population and modeling it within this grid structure. In this case, it is assumed that the intensity of the total distribution of each cluster is proportional to its size. However, this assumption is not always valid. In some situations, cells that belong to the same cluster can have different intensities, e.g. the border cells can present a smaller incidence rate than the central ones. Moreover, a cluster can have a higher incidence of the phenomenon, not because of its size, but due to other factors that influence its disposition, as a spatially referenced covariate. 
	
	This work aims to present a disaggregated model, at cell level, which assumes that the intensity in each cell of a cluster is related to an available covariate value. The proposed model fits rare and grouped populations, disposed over a regular grid, in a Bayesian framework. The key idea of this paper is the improvement of the population estimates through the use of grid cells as analysis units and the incorporation of additional information into the model. Based on this extra information, we also raise an improved sampling process, where different probabilities are assigned to draw the cells, and we can spatially reference the estimates of the cell's count. Introducing additional information seems to be an intuitive idea, provided that the prior knowledge indicates that there is a relationship between the phenomenon occurrence and some covariate.
	
	The paper is organized as follows. In Section \ref{ProposedModel}, the model is introduced, a new sampling procedure is proposed and aspects of inference are discussed. Section \ref{IntroductorySimulationStudy} presents a simulation study for assessing the effectiveness of the proposed model and the model proposed by \cite{Rapley2008}. Section \ref{ComparisonAggregatedModel} shows a comparison of our approach with that of \cite{Rapley2008} through a design-based perspective under different scenarios, as well as a real data application. A simulation study to evaluate the estimation of model parameters under different degrees of rare and clustered populations is presented in Section \ref{SimulationStudy}. Finally, we conclude with a brief discussion about the advantages of our methodology and suggestions for further research in Section \ref{Conclusion}.

	\section{Proposed model for cell counts using covariates}\label{ProposedModel}
	
	Consider a region of interest containing a rare, clustered population, over which a regular grid partitioned into $M$ cells overlaps. Thus, denote the partitioned region by $R=\{c_1,\dots,c_M\}$. A cell is considered nonempty if it contains at least one member of the population and empty otherwise. Let $X \leq M$ be the number of nonempty cells in $R$. Let $P \leq X$ be the number of nonempty networks in $R$, where a network is a set of contiguous non-empty grid cells (\cite{Thompson1990}). Let $Y_i$ be the number of nonempty cells within the network $i$, for $i = 1, \dots, P$, and $\bfY = (Y_1, \dots, Y_P)'$ the vector containing the number of nonempty cells within each nonempty network, so that $X = \sum_{i = 1}^{P} Y_i$. Note that there are $M - X$ empty cells, which are defined as one-sized empty networks, so there are $M - X + P$ networks in $R$. That way, we can extend the $P$-dimensional $\bfY$ vector to a $(M - X + P)$-dimensional vector given by $\bfZ = (\bfY ', \bfone_{M-X}') '$ where $\bfone_{M-X}$ is a vector of ones of dimension $M-X$. Thus, it follows that $Z_i = Y_i$, if the $i$-th network is a nonempty one and $Z_i = 1$, otherwise, for $i = 1, \dots, M - X + P$.
	
	Suppose the phenomenon of interest is related to covariates, which values are available for each one of the cells in $R$. Let $\bfC$ be the set of all nonempty cells of $R$ and $\overline{\bfC}$ the set containing all empty cells of $R$. Let $\eta(c)$ be the count of a given phenomenon of interest in the cell $c$, and $ \bfv_c = (1, v_1(c), \dots, v_k(c))'$ the vector with the $k$ covariates associated with cell $c$, for all $c \in R$. Let $\bfeta$ be the set with the counts for all nonempty cells, that is, $\bfeta=\{ \eta(c) \mid c \in \bfC\}$. 
	
	In order to perform inference about the population total $T =\displaystyle \sum_{c \in \mathbf{C}} \eta(c)$, we must specify the joint distribution of $\{ X, P, \bfY, \bfeta \}$ for the entire population and the sampling mechanism that provides a particular sample of $m$ networks from $M - X + P $ in population. First, we model the nonempty network structure and then, conditional on it, model the count on the nonempty network's cells, similarly to \cite{Rapley2008}'s approach. Since the model applies to nonempty cells, to avoid degeneration problems it is assumed that there is at least one nonempty cell in $R$, so distributions are left truncated at zero. The proposed model can be written as follows:
	\begin{align}\label{Model}
		\eta(c)\mid \bfv_c,\bftheta &\ \sim\ \mbox{truncated Poisson}(\lambda(c)), & \eta(c) & \geq 1,\ c \in \bfC,\nonumber\\ \noalign{\vskip1pt}
		\bfY\mid X,P &\ \sim\ \bfone_P+\mbox{Multinomial}\left(X-P,\frac{1}{P}\bfone_P\right), & Y_i&=1,\dots,X-P,\nonumber\\
		P\mid X,\beta &\ \sim\ \mbox{truncated Binomial}(X,\beta), & P&=1,\dots,X,\nonumber\\ \noalign{\vskip4pt}
		X\mid\alpha &\ \sim\ \mbox{truncated Binomial}(M,\alpha), & X&=1,\dots,M.
	\end{align}
	where $\lambda(c) = \exp\{\bfv_c'\bftheta\}$, $\bftheta = (\theta_0, \theta_1, \dots, \theta_k)'$ represents the regression coefficients vector associated with $\bfv_c$. Note that the $M - X$ empty cells have their respective counts equal to zero, that is, $\eta(c) = 0 $, for all $c \in \overline{\bfC}$.
	
	\subsection{Model inference}\label{ModelInference}
	
	The sampling procedure entails observing $Y_i$ for the networks $\{ i_1, \dots i_m\}$ and the counts $\eta(c)$ for its respective cells. Since adaptive cluster sampling procedure depends on the population structure, it is characterized as an informative sampling design and the probability of selecting the sample $s = \{i_1, \dots, i_m \}$ of $m$ networks, $[s\mid X,P,\bfY]$, should be incorporated into the model likelihood function. Set the subscript `$s$' to identify the observed component and $\sbar$ to the unobserved component, and define $\bfY = (\bfY_s', \bfY_{\sbar}')'$, $X = X_s + X_{\sbar}$ and $P = P_s + P_{\sbar}$ to distinguish between observed and unobserved quantities. Let $\bfC_s$ be the set of the sample's nonempty cells, i.e., the cells that compose the networks with sizes $\bfY_s$; and $\bfC_{\sbar}$ be the set of the out-of-sample nonempty cells, i.e., the cells that compose the non-sampled networks with sizes $\bfY_{\sbar}$. Thus, define $\bfeta = (\bfeta_s', \bfeta_{\sbar}')'$, where $\bfeta_s = \{\eta(c) \mid c \in \bfC_s\}$ and $\bfeta_{\sbar} = \{\eta(c) \mid c \in \bfC_{\sbar}\}$. A natural predictor of the population total $T$ is given by:
	$$
	T = \sum_{c \in \bfC_s}\eta(c) + \sum_{c \in \bfC_{\sbar}}\hat{\eta}(c),
	$$
	where $\hat{\eta}(c)$ is the estimated count of the cell $c$, for $c \in \bfC_{\sbar}$.
	
	Following the Bayesian paradigm, it is also assumed independent priors for the unknown parameters $\bftheta$, $\alpha$ and $\beta$ and their marginal prior distributions are denoted, respectively, by $[\bftheta]$, $[\alpha]$ and $[\beta]$. Let $[\bftheta]$ be a non-informative prior with a zero-mean vector and covariance matrix $\sigma^2_\theta\bfI_{k+1}$, where $\bfI_{k+1}$ denotes the $(k+1)$-dimensional identity matrix. For $\alpha$ we assumed a Beta$(a_\alpha, b_\alpha)$ distribution with $a_\alpha = 3$ and $b_\alpha = 15$, and for $\beta$ a Beta$(a_\beta, b_\beta)$ distribution with $a_\beta = 1$ and $b_\beta = 9$. The prior distributions of $\alpha$ and $\beta$ are chosen to reflect the fact that $\alpha$ and $\beta$ are necessarily small in a rare and clustered population, as considered in \cite{Rapley2008}. In this case, the objective is not only to estimate the parameters of the model based on a sample, but also to make predictions of the unobserved parts.
	
	The joint distribution of all the quantities in the model is:
	\begin{eqnarray}\label{posterior}
	&&[\bfeta,\bfY,P,X,\bftheta,\beta,\alpha]\nonumber\\ [5pt]
	&=& [s\mid X,P,\bfY][\bfeta\mid \bftheta][\bfY\mid X, P][P\mid X,\beta][X\mid\alpha][\bftheta][\alpha][\beta]\nonumber\\ [5pt]
	&\propto&[s\mid X,P,\bfY]\times\prod_{c \in \bfC}\frac{\exp\{-\exp\{\bfv_c'\bftheta\}+\eta(c)\bfv_c'\bftheta\}}{\eta(c)!( 1-\exp\{-\exp\{\bfv_c'\bftheta\}\}) }\nonumber\\[8pt]
	&&\times(x-p)!\prod_{i=1}^p\frac{1}{(y_i-1)!}\left(\frac{1}{p}\right)^{y_i-1}\times\left(\begin{array}{c}x\\p\end{array}\right) \frac{\beta^{p}(1-\beta)^{x-p}}{1-(1-\beta)^x}\times\left(\begin{array}{c}M\\x\end{array}\right)\frac{\alpha^{x}(1-\alpha)^{M-x}}{1-(1-\alpha)^M}\nonumber\\ [8pt]
	&&\times\exp\left\{-\frac{1}{2\sigma^2_\theta}\bftheta'\bftheta\right\}\times \alpha^{a_\alpha-1}(1-\alpha)^{b_\alpha-1}\times \beta^{a_\beta-1}(1-\beta)^{b_\beta-1}.
	\end{eqnarray}
	
	We perform inference via MCMC to obtain samples from the resulting posterior distribution. The full conditional posterior distributions and the methods adopted to sample from each of them are detailed in Appendix \ref{AppendixA}. In comparison with the sampling procedure proposed by \cite{Salehi1997}, our improved sampling process leads us to draw a greater number of networks, providing samples that may include all networks from $R$ (see details in Subsection \ref{SamplingProcedure}). Thus, our proposal distribution, different from \cite{Rapley2008}'s approach, may lead to none out-of-sample nonempty cells and, consequently, none out-of-sample networks (see details in Appendix \ref{AppendixA}). The estimation procedure consists of the following steps: \vspace{-0.2cm}
	\begin{enumerate}
		\item[(1)] Initialize the counter $j=2$ and set initial values for the parameters and quantities of the model: $\bftheta^{(1)}$, $\alpha^{(1)}$, $\beta^{(1)}$, $X_{\sbar}^{(1)}$, $P_{\sbar}^{(1)}$, $\bfY_{\sbar}^{(1)}$ and $\bfeta_{\sbar}^{(1)}$;\vspace{-0.1cm}
		\item[(2)] Update the model parameters $\bftheta$, $\alpha$ and $\beta$ from the conditional distributions:\vspace{-0.3cm}
		$$[\bftheta \mid \alpha^{(j-1)}, \beta^{(j-1)}, X^{(j-1)},P^{(j-1)},\bfY^{(j-1)},\bfeta^{(j-1)}],\vspace{-0.2cm}$$
		$$[\alpha \mid \bftheta^{(j)}, \beta^{(j-1)}, X^{(j-1)},P^{(j-1)},\bfY^{(j-1)},\bfeta^{(j-1)}],$$
		$$[\beta \mid\bftheta^{(j)}, \alpha^{(j)}, X^{(j-1)},P^{(j-1)},\bfY^{(j-1)},\bfeta^{(j-1)}],\vspace{-0.2cm}$$
		described in  Appendix \ref{AppendixA};\vspace{-0.1cm}
		\item[(3)] Generate the non-sampled quantities $X_{\sbar}$, $P_{\sbar}$ and $\bfY_{\sbar}$ according to the proposal distribution described in Subsection \ref{FullConditionalQuantities};\vspace{-0.2cm}
		\item[(4)] Allocate the $P_{\sbar}$ networks of $\bfY_{\sbar}$ according to the allocating procedure described in Subsection \ref{AllocatingProcedure};\vspace{-0.2cm}
		\item[(5)] Generate $\bfeta_{\sbar}$ and jointly update $X_{\sbar}$, $P_{\sbar}$, $\bfY_{\sbar}$ and $\bfeta_{\sbar}$ from the conditional distribution:\vspace{-0.3cm}
		$$[X_{\sbar},P_{\sbar},\bfY_{\sbar},\bfeta_{\sbar}\mid \bftheta^{(j)},\alpha^{(j)},\beta^{(j)}, X_s,P_s,\bfY_s,\bfeta_s];\vspace{-0.4cm}$$
		\item[(6)] Increment the counter $j$ to $j+1$ and interate from (2).
		\vspace{-0.2cm}
	\end{enumerate}
	
	Note that the regression coefficients $\bftheta$ are updated on step (2) based only on the sample information. Moreover, from them, we can easily obtain the Poisson distribution's intensity $\lambda(c) = \exp\{\bfv_c'\bftheta\}$ for all non-sampled cell $c$ of $R$, which is used later to estimate $\eta(c)$ for all nonempty and non-sampled cell $c$ of $R$. Let $\bflambda$ be the set of intensities assigned to all cells of $R$. Then, after generating the non-sampled quantities $X_{\sbar}$, $P_{\sbar}$ and $\bfY_{\sbar}$, all that remains is to find out which cells form each of these $P_{\sbar}$ networks on step (4), according to the allocating procedure presented in Subsection \ref{AllocatingProcedure}.
	
	\subsubsection{Allocating procedure}\label{AllocatingProcedure}
	
	Determining the cells that compose the out-of-sample nonempty networks is a crucial step in the proposed model estimation since the resulting allocation directly impacts: the cells that compose $\bfC_{\sbar}$ and the estimated value of $\bfeta_{\sbar}$. Each one of the generated out-of-sample networks are allocated sequentially, according to its size: the bigger networks are allocated first and the smaller ones later. It is assumed that the bigger the size of the network, the higher its cells' intensity values. Note that the cells that compose the set of the out-of-sample cells, $\bfC_{\sbar}$, must not be part of the set of sampled cells, $\bfC_{s}$, nor of the sampled nonempty networks' borders (if it happens, we would be able to modify a network previously sampled).
	
	\begin{figure}[ht!]
		\centering
		\begin{tabular}{@{}c@{}} \subfigure[]{
				\includegraphics[width=4.7cm]{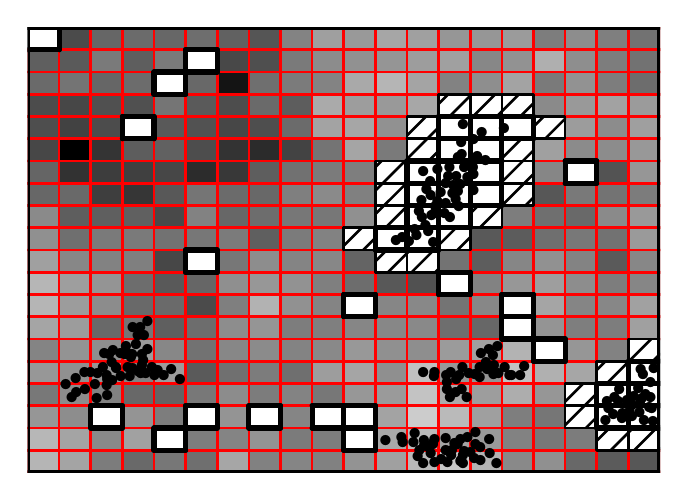}}\end{tabular}
		$\rightarrow$
		\begin{tabular}{@{}c@{}} \subfigure[]{
				\includegraphics[width=4.7cm]{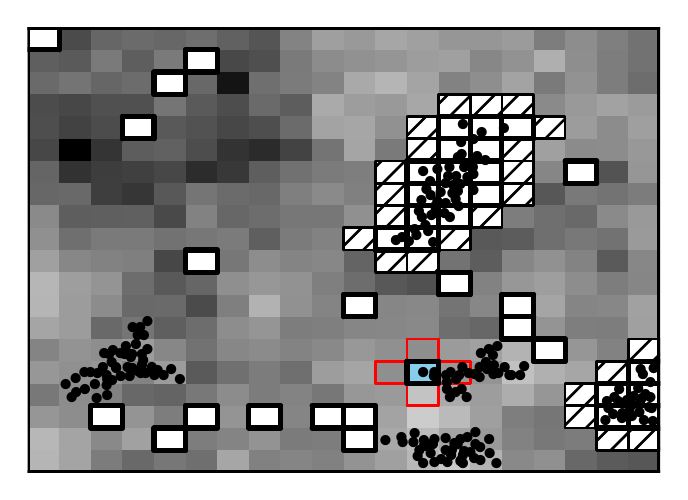}}\end{tabular}
		$\rightarrow$
		\begin{tabular}{@{}c@{}} \subfigure[]{
				\includegraphics[width=4.7cm]{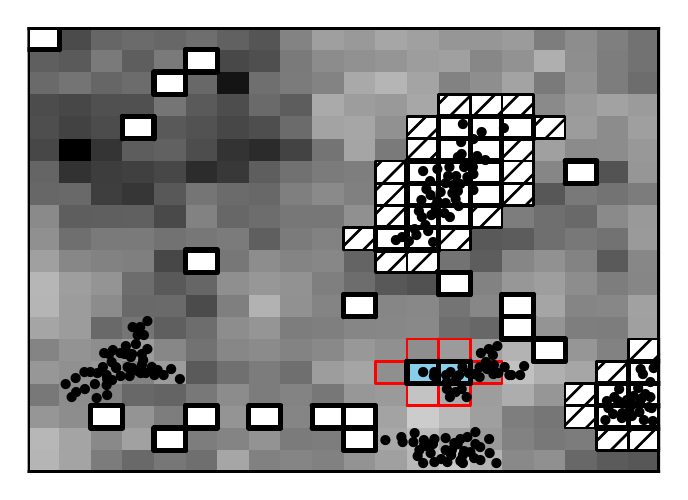}}\end{tabular}
		\begin{tabular}{@{}c@{}} \subfigure[]{
				\includegraphics[width=4.7cm]{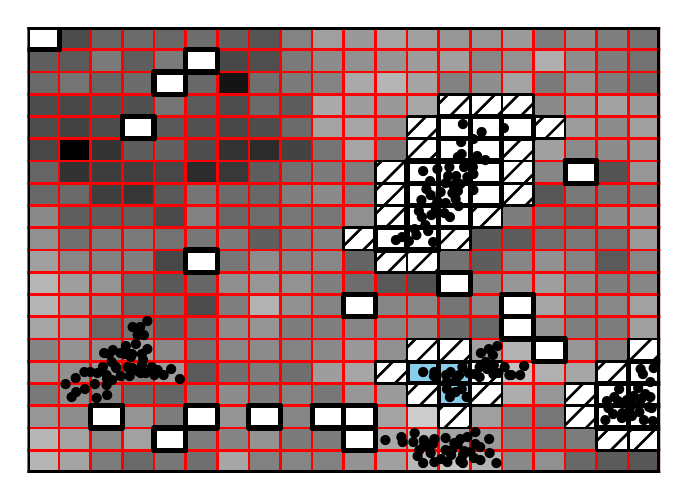}}\end{tabular}
		$\rightarrow$
		\begin{tabular}{@{}c@{}} \subfigure[]{
				\includegraphics[width=4.7cm]{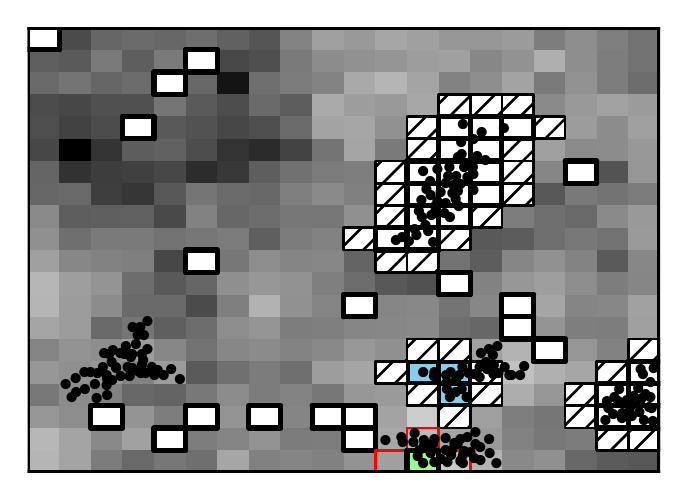}}\end{tabular}
		$\rightarrow$
		\begin{tabular}{@{}c@{}} \subfigure[]{
				\includegraphics[width=4.7cm]{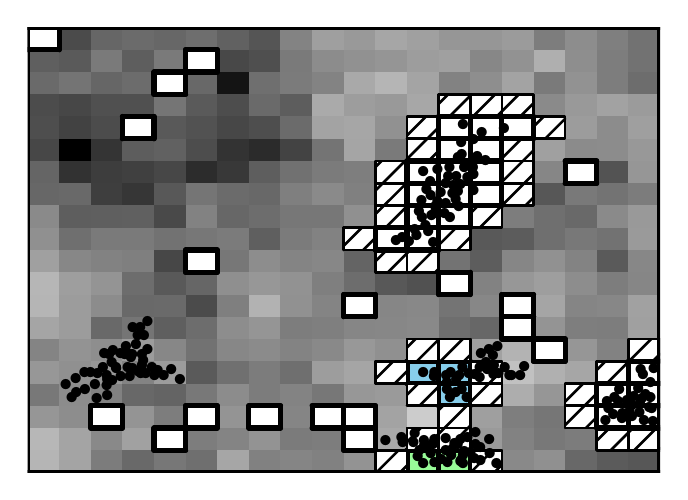}}\end{tabular}\vspace{-0.2cm}
		\caption{\small Allocation method illustration of two out-of-sample networks of sizes 3 and 2, based on weights $\bflambda$ (gray background). The lighter the cells' color, the higher the $\lambda$ value (intensity) of that cell. The white cells with bold borders, whose weights are equal to zero, are the sampled networks' cells and the hatched ones correspond to the nonempty networks' border cells. The red borders surround cells that can be drawn in each stage of the procedure. The cells that compose the first and second allocated networks are blue-painted and green-painted, respectively. The example proceeds as follows: draw one of the red-surrounded cells of Panel (a). The sampled cell is blue-indicated in Panel (b) and only its neighbors can be sorted to keep building this network.  In Panel (c) and (d) the allocation of the 3-sized network is finished, and we can draw any cell with the red border to start the allocation of the 2-sized network. Panels (e) and (f) present the cells chosen to compose this network in green.}
		\label{AllocatingProcess}
	\end{figure}\vspace{-0.2cm}
	
	The allocating procedure aims to draw the cells that compose each generated out-of-sample network according to determined weights. In this case, we will use the set of intensities $\bflambda$, although one could sample the cells based on other practical weights. The $\bfC_{s}$ cells' and visited borders' weights $\bflambda$ are admitted to be zero. An example of this procedure is illustrated in Figure \ref{AllocatingProcess}. The allocating method of a network of size $Y$ proceeds as follows: drawn an available cell $c$ with probability proportional to the weights $\bflambda$ and, if $Y > 1$, drawn another cell from the neighbors of that cell and continue to drawn another neighbors' cells until we obtain a set of $Y$ contiguous nonempty grid cells surrounded by empty grid cells. Then remove this network from the population, select one of the remaining grid cells with probability proportional to the weights $\bflambda$ and proceed in this way until we have allocated all the $P_{\sbar}$ networks. Note that the cells that were not chosen to be part of $\bfC_{\sbar}$ are assumed to be empty.

	
	
	\subsection{Sampling procedure}\label{SamplingProcedure}
	
	A variation of the sampling procedure proposed by \cite{Salehi1997} is proposed here to improve the sampling process, aiming to sample more nonempty networks. Let $\bfpi$ be the set of sampling weights assigned to all cells of $R$ and $\pi(c)$ the weight for a given cell $c$. The procedure consists of sampling a grid cell from the set of $M$ grid cells with probability proportional to the weights $\bfpi$ and, if it is nonempty, the entire network containing the selected grid cell. After removing this network from the population, a new one cell is selected from the remaining set of grid cells and the method proceed in this way until we have selected $m$ networks in the sample. Note that a nonempty network is surrounded by empty cells that make up its border and can be resampled.
	
	\begin{figure}[ht!]
		\centering
		\begin{tabular}{@{}c@{}} \subfigure[Population and constant weights]{
				\includegraphics[width=6.8cm]{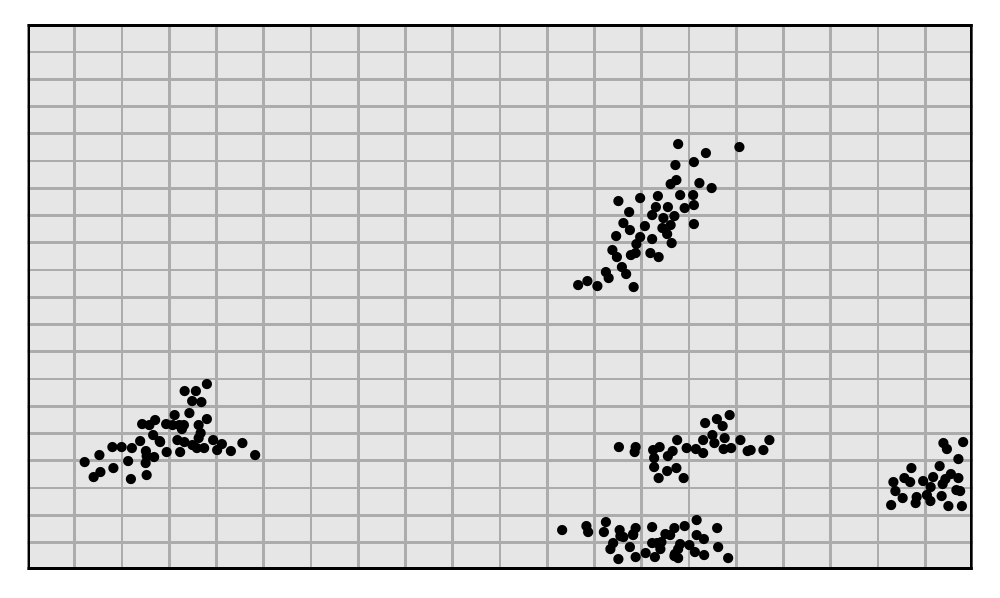}}\end{tabular}
		{$\xrightarrow{1^{st}\mbox{ stage}}$}
		\begin{tabular}{@{}c@{}} \subfigure[First sample: $m_1$ networks]{
				\includegraphics[width=6.8cm]{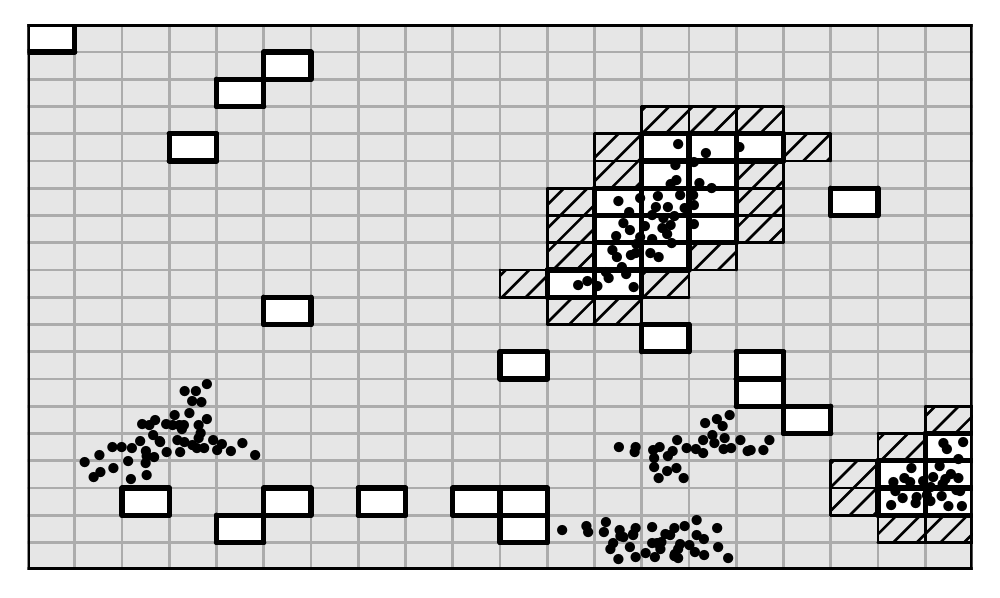}}\end{tabular}
		\begin{tabular}{@{}c@{}} \subfigure[First sample and weights $\bfomega$]{
				\includegraphics[width=6.8cm]{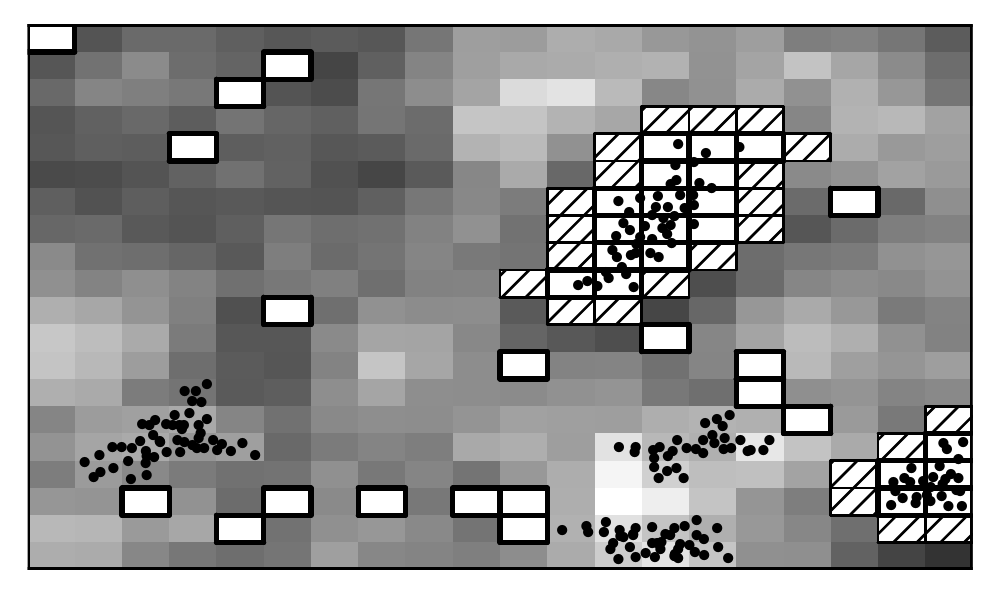}}\end{tabular}
		{$\xrightarrow{2^{nd}\mbox{ stage}}$}
		\begin{tabular}{@{}c@{}} \subfigure[Final sample: $m1 + m_2$ networks]{
				\includegraphics[width=6.8cm]{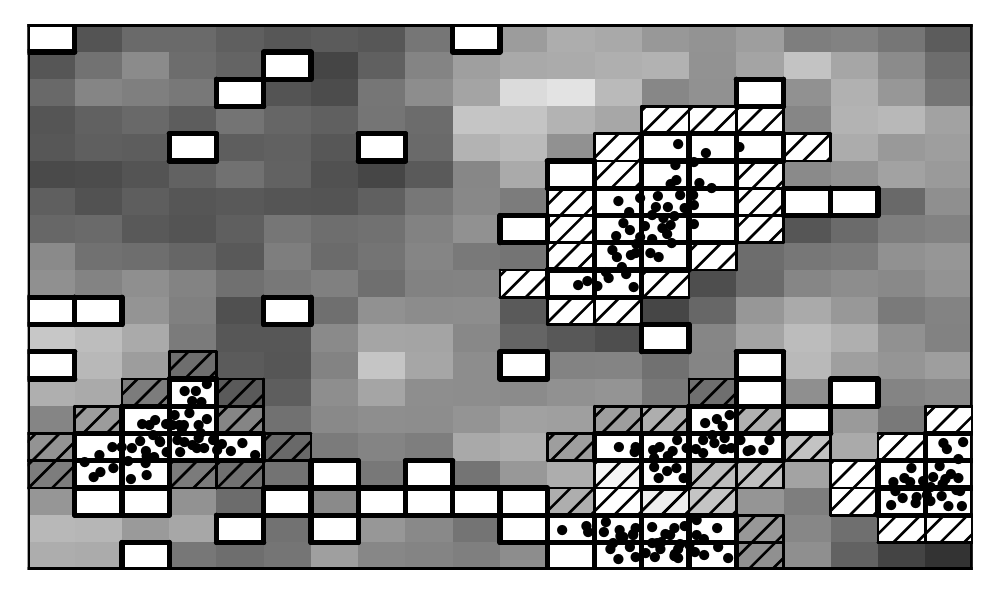}}\end{tabular}
		\caption{\small Proposed sampling procedure illustration of a population (points) distributed in a region with $M=400$ cells and weights (gray background) used in this scheme. The lighter the cells' color, the higher the weight of that cell. The white cells, whose weights are equal to zero, with bold borders are the sampled networks' cells and hatched cells correspond to the nonempty networks' border cells. Panels (a) and (b) present the same grayscale, since all cells have constant weight in the first stage; and Panels (c) and (d) show different shades of gray due to the second stage's different weights.}
		\label{SamplingProcedureIlustration}\vspace{-0.3cm}
	\end{figure}
	
	The sampling process improvement proposed in this paper, illustrated in Figure \ref{SamplingProcedureIlustration}, is divided into two stages and is based on weights that are used to draw the sample. In the first stage, $m_1$ networks are selected considering grid cells with equal weights, i.e. $\pi(c)$ is constant for all $c \in R$. The sampling procedure continues until all nonempty cells in the neighborhood are observed and stop when empty units are visited. Thus, the networks are selected with probability proportional to their size. Note that, during this process, although the border cells are visited, they are not added to the sample. Based on the fit of the proposed model in equation (\ref{Model}) to this first sample with $m_1$ networks, we obtain the vector of weights $\bfomega$ for all non-sampled cells of $R$, which are used to collect the second sample. Let $\omega(c)$ be the weight defined by the posterior mean of $\eta(c)$, for each cell $c \in R$. Note that the higher the posterior mean of a cell count, the more chances of selecting that cell. Due to the inference process, the weights $\bfomega$ associated with the border cells are assigned to be zero. Since the first sample of the network's cells must not be drawn in the second sampling stage, the weights associated with these cells are assumed to be zero too. Then, a second sample of $m_2$ networks is drawn with probability proportional to the weights $\bfomega$. Hence, the final sample will be given by $s = s_1\cup s_2 = \{ i_1, \dots, i_{m_1}, i_{m_1 + 1}, \dots, i_{m_1 + m_2} \}$, with size $m = m_1 + m_2$.
	
	
	To motivate the notation for the probability of selecting a given sample, consider a population consisting of networks of size $\bfZ$ from which we obtain the ordered sample $s = \{i_1, \dots, i_m \}$. The probability of selecting the $j$-th network of the sample, that is a network of size $Z_{i_j}$, is given by the sum of probabilities of selecting each unselected network of size $Z_{i_j}$ after $j-1$ networks have been observed, since networks with the same size are considered alike. Thus, the probability of selecting a network in the sample depends on its size $Z_i$, which is only observed for the sampled networks after their selection in the sample.
	
	Let $\bfc_{j} $ be the set of sampled cells in the $j$-th draw. Thus, $\bfc_{j}$ is composed of the drawn grid cell and, if it is nonempty, $\bfc_{j}$ contains the entire network containing the selected grid cell. Let $\bfG_{i_j, j}$ be the set of cells that compose unselected networks of size $Z_{i_j}$ after $j-1$ networks have been selected. Thus, in general, the probability of selecting the sample $s = \{i_1, \dots, i_m \}$ of $m$ networks is given by:
	\begin{eqnarray}\label{SelectionProbability}
	[s\mid X,P,\bfY]\ =\ \prod_{j=1}^{m} \sum_{g\in\bfG_{i_j,j}} \frac{\displaystyle \pi(g)} {\displaystyle \sum_{r \in R}\pi(r) - \sum_{k=0}^{j-1}\sum_{c\in\bfc_{k}}\pi(c)},
	\end{eqnarray}
	where $\pi(c)$ represents the weight of the cell $c$ and is:
	$$
	\pi(c) = \left\{ \begin{array}{ll}
	\mbox{constant}, & \mbox{if $c \in s_1$};\\
	\omega(c), & \mbox{if $c \in s_2$}.\end{array} \right. 
	$$
	Note that in equation (\ref{SelectionProbability}), the index $j$ represents $j$-th draw, so $c \in s_1$ for $j=1, \dots, m_1$, and $c \in s_2$ for $j = m_1+1, \dots, m$. When the proposed model in equation (\ref{Model}) is fitted to the first sample (to obtain the weights $\bfomega$), the weights $\pi(c)$ are constant and the probability given in expression (\ref{SelectionProbability}) matches with the probability of selecting a sample $s$ given in \cite{Rapley2008}. On the other hand, differently from \cite{Rapley2008}, the probability of selecting a given sample $s$ does not depend directly on the quantities of the model, but on the weights of the networks' cells.
	
	The cells that compose each non-sampled network are defined from their allocation process (described in Subsection \ref{AllocatingProcedure}), which directly impacts the weights $\bfomega$ used to collect the second sample. Therefore, the proposed model must properly determine the cells that compose the out-of-sample nonempty networks. If a cell is part of $\bfC_{\sbar}$ in a large number of MCMC iterations, this cell tends to be a nonempty cell of $R$ and the associated posterior mean will be high, while, if a cell does not compose $\bfC_{\sbar}$ in a large number of MCMC iterations, this cell tends to be an empty cell of $R$. It is expected that this novel sampling method based on weights will lead us to a more efficient draw of networks, as we are assigning higher chances to the cells where the phenomenon of interest is expected to be found and avoiding sampling in areas where the expected intensity of the phenomenon’s occurrence is low.
	
	The proposed sampling methodology consists of the following steps:\vspace{-0.2cm}
	\begin{enumerate}
		\item[(1)] Consider a region $R$ containing a rare, clustered population, partitioned into $M$ cells and drawn an adaptive cluster sample of $m_1$ networks, which is equivalent to drawing a sample of $m_1$ networks with probability proportional to their sizes, i.e., the elements of the vector of probabilities $\bfpi$ are constant;\vspace{-0.2cm}
		\item[(2)] Fit the proposed model in equation (\ref{Model}) to this first sample to obtain the posterior mean of the cells' counts $\eta(c)$, given in the vector $\bfomega$, which will be used as weights to collect the second sample;\vspace{-0.2cm}
		\item[(3)] Since the first sample of the network's cells must not be drawn in the second sampling stage, set the weights associated with the cells of the first sample as zero, as well as, the non-empty networks' border cells;\vspace{-0.2cm}
		\item[(4)] From the remaining cells of $R$, drawn $m_2$ networks with probability proportional to the weights $\bfomega$;\vspace{-0.2cm}
		\item[(5)] Finally, fit the proposed model in equation (\ref{Model}) to the final sample of size $m=m_1+m_2$ to estimate the population total.
	\end{enumerate}
	
	\subsubsection{Border-sampling procedure}\label{BorderSamplingProcedure}
	
	Through the proposed sampling method, we survey a drawn grid cell and, if it is nonempty, the entire network containing the selected grid cell. It is important to remark that nonempty networks are surrounded by empty cells that compose its border, which are not removed from $R$ unless they are drawn as an empty network. Thus, a surveyed border cell can be drawn later, although we know that it is empty.
	
	To avoid surveying the same border cell twice, we propose an alternative sampling method, given as follows: draw a grid cell from $R$ with probability proportional to the weights $\bfpi$, survey that grid cell and, if it is nonempty, survey the entire network containing the selected cell. After removing this network and its border from the population, select a new one cell from the remaining set of grid cells and proceed in this way until we have selected $m$ networks in the sample. In practice, proceeding this way is equivalent to surveying clusters instead networks, though the final sample structure is the same as before but containing the border cells' information. Note that the only change in this method is that the border cells can not be re-sorted, in comparison with the method previously presented in Subsection \ref{SamplingProcedure}.
	
	The inference procedure is the same as Subsection \ref{ModelInference} except for the joint distribution of all the quantities in the model (\ref{posterior}) since the probability of selecting a given sample $s$ has changed. Let $\bfc_{j} $ be the set of sampled cells in the $j$-th draw. In this case, $\bfc_{j}$ is composed of the drawn grid cell and, if it is nonempty, the entire network containing the selected grid cell plus its border. This subtle change in the sets $\bfc_{j}$, for $j = 1, \dots, m$, incorporates the sampling modification. Thus, the expression of the probability of selecting the sample $s$ of $m$ networks is the same as in (\ref{SelectionProbability}) except for the definition of the set $\bfc_{j}$.
	
	
	
	\section{A preliminary simulation study}\label{IntroductorySimulationStudy}
	
	In order to assess the effectiveness of each methodology, we compared the results of our approach to those obtained in \cite{Rapley2008}. In this section, we will refer to the model of \cite{Rapley2008} as the ‘network-sampling aggregated model' (MAN) since the sampling procedure used in this methodology allows the networks' border cells to be re-sorted. Analogously, the model proposed in Section \ref{ProposedModel} considering the sampling procedure firstly presented in Subsection \ref{SamplingProcedure} will be referred to as the ‘network-sampling disaggregated model' (MDN), and the one considering the methodology proposed in Subsection \ref{BorderSamplingProcedure} as the ‘cluster-sampling disaggregated model' (MDC), since the sampling procedure does not allow the networks' border cells to be re-sorted, i.e., the entire cluster is sampled.
	
	\begin{figure}[ht]
		\begin{center}
			\includegraphics[width=10cm]{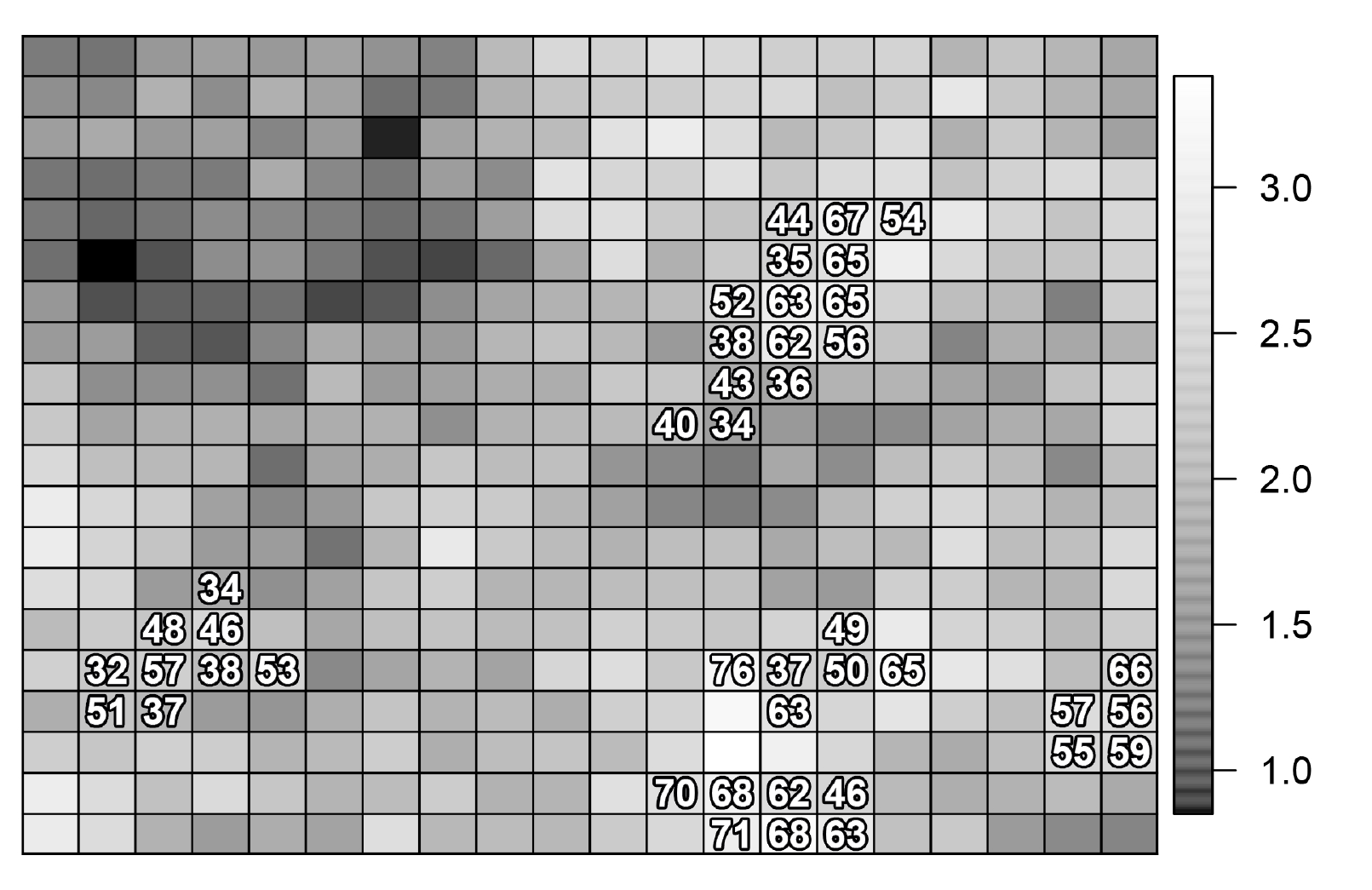}\vspace{-0.3cm}
			\caption{\small Values of the generated covariate (gray background) and counts for each nonempty cell in a grid with $M = 400$ cells. }\label{SimulatedData}
		\end{center}\vspace{-0.7cm}
	\end{figure}
	
	The population used here was generated based on the disaggregated model presented in equation (\ref{Model}) according to the fixed parameters $(\alpha, \beta) = (0.1, 0.1)$ and $(\theta_0, \theta_1)=(2.7, 0.5)$. The fictional covariate was simulated from a gaussian process. Figure \ref{SimulatedData} shows the generated population and covariate in a grid with $M=400$ cells. Note that these counts are sparse and clustered, motivating the use of adaptive sampling. \vspace{-0.2cm}
	
	
	\begin{figure}[ht!]
		\begin{center}
			\includegraphics[width=15.5cm]{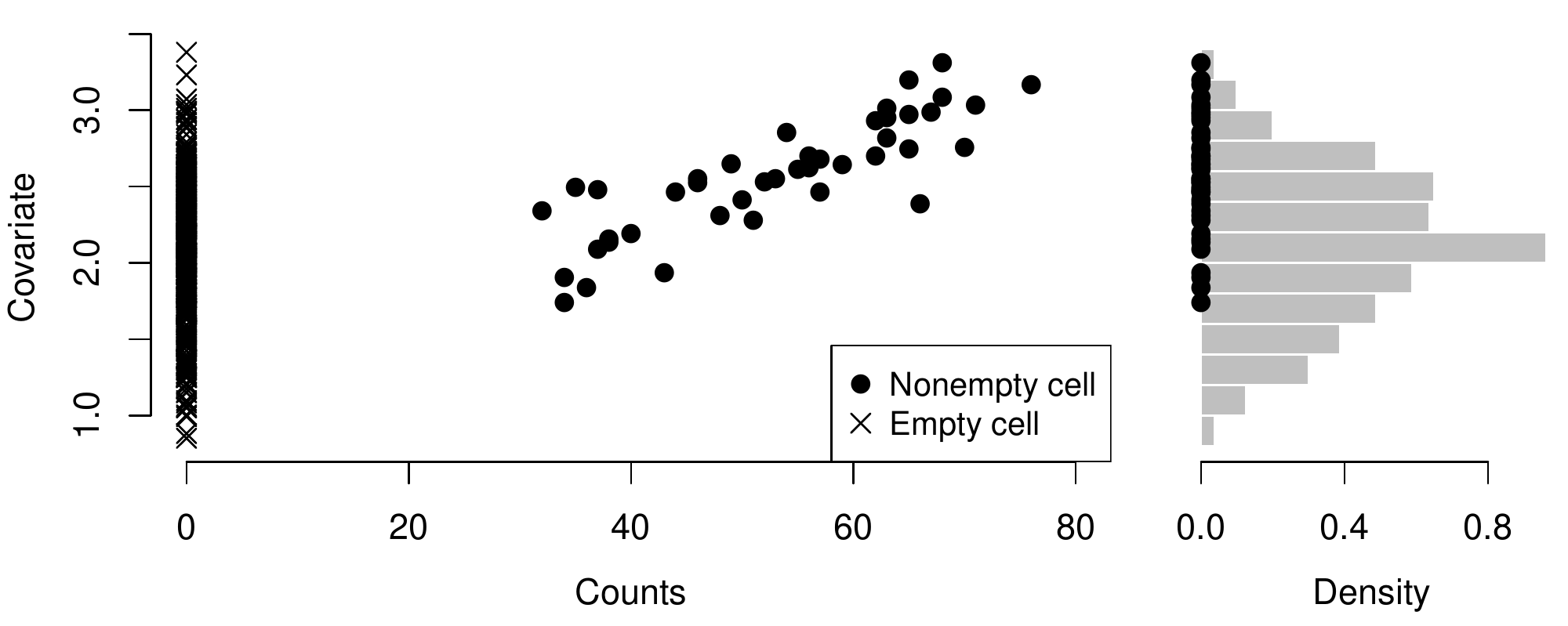}\vspace{-0.3cm}
			\caption{\small Plot of the generated covariate versus the population counts and covariate's histogram, where the black points in the histogram represent the nonempty cells.} \label{SimulatedDataCovariate}
		\end{center}\vspace{-0.7cm}
	\end{figure}
	
	We can observe from Figure \ref{SimulatedDataCovariate} that the higher the generated covariate value, the higher the associated count of nonempty cells. Moreover, the simulated interest event is associated with higher values of the covariate since there is no occurrence of that in areas with lower covariate values.
	
	The study consists of drawing 100 nonempty samples of $m=40$ networks of the population according to each method. The proposed sampling methodologies are divided into two stages, where we sample $m_1 = 20$ networks randomly and $m_2 = 20$ networks based on weights $\bfomega$, according to the proposed sampling methodology. Note that we are using an initial sample of size 50\% of $m$ to obtain these weights and one can adjust this percentage (see Subsection \ref{ModelBasedExperiment}). In the study performed in this section, we are omitting samples that consist of only empty networks since our proposed models require at least one nonempty network in the first sample to adjust the weights $\bfomega$ properly.
	
	The MCMC algorithm was implemented in the \texttt{R} programming language, v. 3.6.1 (\cite{R2019}). For each sample and fitted model, we ran two parallel chains starting from different initial values, let each chain run for 40,100 iterations, discarded the first 100 as burn-in, and stored every 20th iteration to obtain 2,000 independent samples. We used the diagnostic tools available in the package \texttt{CODA} (\cite{Coda2006}) to check convergence of the chains. Convergence results of Subsection \ref{DesignBasedExperiment} are available in Appendix \ref{AppendixB}.
	
	A summary comparison of the population total estimators using the relative square root of the mean square error (RRMSE), relative absolute error (RAE), relative bias (RB), relative width (RW) and the empirical coverage of the 95\% credibility intervals measured in percentages (Cov.) are presented in Table \ref{IntroStudySummary}. These results are obtained considering all 100 samples generated according to each model.
	
	\begin{table}[ht]
		\centering 
		\begin{tabular}{c|ccccc|}
			\cline{2-6}
			\multicolumn{1}{c|}{}& RRMSE & RAE & RB & RW & Cov.\\ \hline 
			\multicolumn{1}{|c|}{MAN} & 0.296 &  0.236 &  -0.009 & 0.879 & 95.88\\ 
			\multicolumn{1}{|c|}{MDN} & 0.265 & 0.209 & -0.022 & 0.756 & 97.94\\ 
			\multicolumn{1}{|c|}{MDC} & 0.283 & 0.217 &  0.014 & 0.878 & 100.00\\ \hline 
		\end{tabular}
		\caption{\small Summary measurements of the point and 95\% credibility interval estimates of the population total $T$, obtained by fitting MAN, MDN and MDC models under 100 samples generated according to each model.}\label{IntroStudySummary}\vspace{-0.2cm}
	\end{table}
	
	Table \ref{IntroStudySummary} shows that the Bayes estimator produced by MDN fit has a smaller RRMSE and RAE than MAN and MDC's estimators. The model MAN produced the smallest RB, but it seems to be less efficient than the other models according to the other error measurements. Although the MDC produces 95\% credibility intervals with higher coverage percentages than the others, its width (RW) is not the smallest one. By this way, the proposed model MDN appears to be more efficient when applied to these artificial samples.\vspace{0.1cm}
	
	\begin{figure}[ht!]
		\begin{center}
			\includegraphics[width=15.5cm]{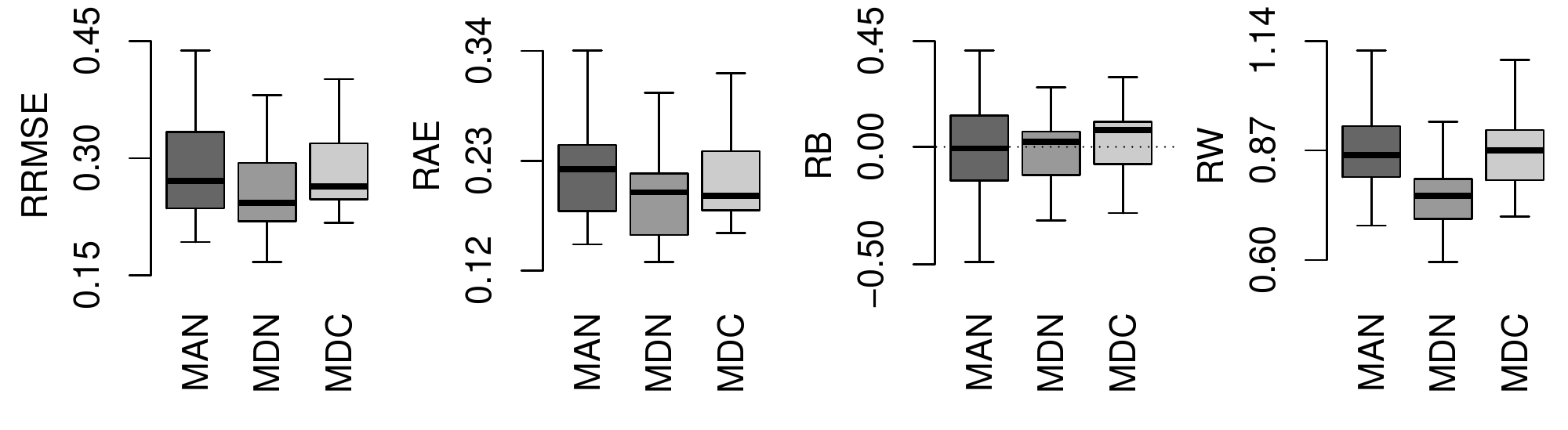}\vspace{-0.3cm}
			\caption{\small Boxplots with measurements of the point and 95\% credibility interval estimates for $T$ over 100 simulations obtained for the fits of MAN, MDN and MDC models.}\label{Boxplots3Models}
		\end{center}\vspace{-0.8cm}
	\end{figure}
	
	In a similar way, Figure \ref{Boxplots3Models} shows the boxplots of the RRMSE, RAE, RB and RW of the Bayes estimators obtained when fitting each model, based on all 100 samples. Here again, we see that the RRMSE, RAE and RW obtained for MDN model are lower than the others'. Note that the RB distributions are quite similar although MDN one has a smaller variability than the others'.
	
	\begin{figure}[ht!]
		\begin{center}
			\includegraphics[width=15.5cm]{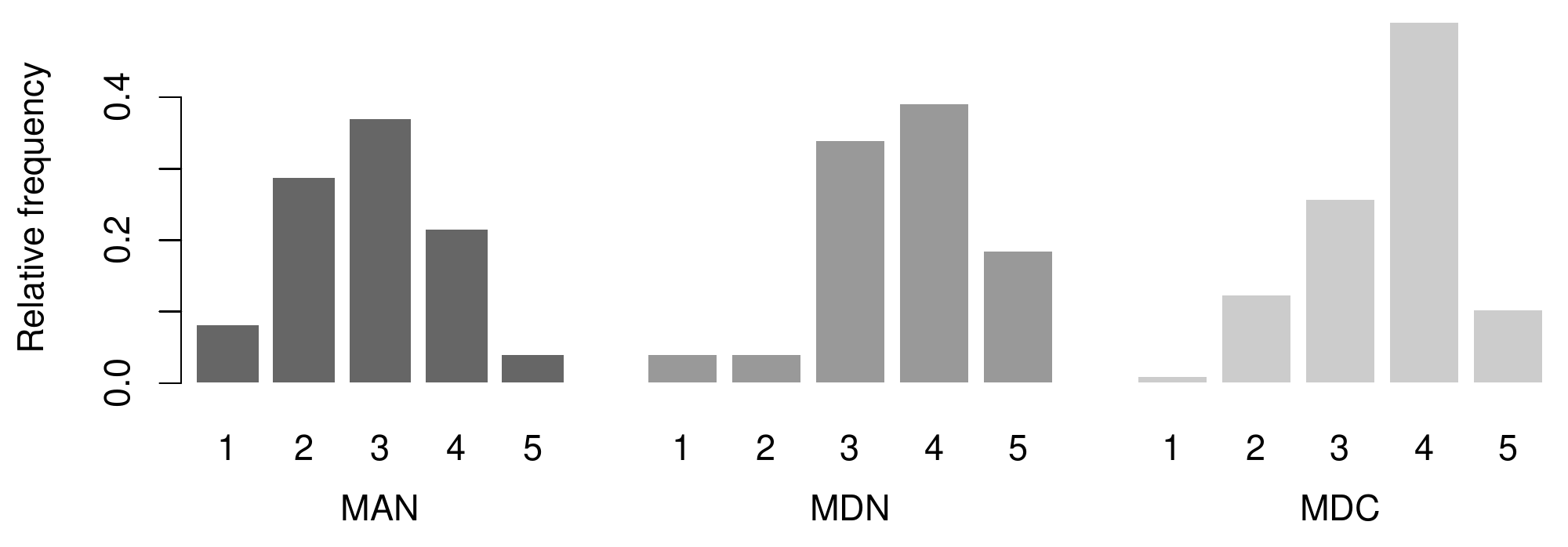}\vspace{-0.3cm}
			\caption{\small Relative frequency of the number of networks sampled over 100 simulations obtained for the fits of MAN, MDN and MDC models.}\label{Networks3Models}
		\end{center}\vspace{-0.8cm}
	\end{figure}
	Finally, we present the barplots with the relative frequency of the number of networks sampled from 100 fits of each model in Figure \ref{Networks3Models}. Note that the sampling procedure of the proposed disaggregated models tend to sample more networks than the aggregated one. In particular, MDN sampling procedure provides us more samples containing the whole population than the others.
	
	Since among the proposed methodologies, the MDN model (which allows the border cells to be re-sorted) yielded better results than the MDC model, we will focus on studying the properties of the MDN model from now on, as well as a more extensive comparison with the aggregated model and an application to a real data.
	
	\section{Comparison with the aggregated model}\label{ComparisonAggregatedModel}
	
	
	To assess the effectiveness of our proposed methodology, we compared the results of our approach considering the MDN model of Section \ref{IntroductorySimulationStudy}, which will be simply referred to as disaggregated model (MD), to those obtained in \cite{Rapley2008}, called aggregated model (MA). The first comparison consists of a design-based experiment, where the numbers of networks $m_1$ and $m_2$ selected are studied, and the second one is a real experiment with an African Buffalo population in an area of East Africa.
	
	\subsection{A design-based experiment evaluating the sample fraction}\label{ModelBasedExperiment}
	
	The purpose of this simulation study is to compare the performance of the aggregated and disaggregated models when the population is generated according to the disaggregated model, and study how the choice of the numbers of networks $m_1$ and $m_2$ selected, respectively, in the first and second sampling stages, affect the population total estimates. We considered twelve scenarios to evaluate how the sample size $m$ and the numbers $m_1$ and $m_2$ of networks selected in the first and second sampling stages, respectively, affect the population total estimates under the disaggregated model. We fixed the total sample size $m \in \{30, 40, 50\}$ and the percentage of the $m$ networks to be sampled in the first sampling stage at $\{35\%, 50\%, 65\%, 80\%\}$, i.e., the numbers $m_1$ and $m_2$ depend on these percentages. We used the same population generated in the simulation study presented in Section \ref{IntroductorySimulationStudy}, which is distributed in a region with $M=400$ cells, drawn 500 samples according to each scenario and methodology, and fitted both models to evaluate its performances. Note that, the aggregated model's sampling methodology considers only one sample of size $m$, collected as in the first sampling stage of the proposed methodology. Table \ref{ModelBasedTableSizes} shows the values of the sample size $m$ and the respective $m_1$ and $m_2$ of networks selected in the first and second sampling stages, according to each fixed percentage.
	
	\begin{table}[ht!]
		\centering
		\begin{tabular}{c|cc|cc|cc|cc|}
			\cline{2-9}
			& \multicolumn{2}{c|}{35\%} & \multicolumn{2}{c|}{50\%} & \multicolumn{2}{c|}{65\%} & \multicolumn{2}{c|}{80\%} \\ \hline
			\multicolumn{1}{|c|}{$m$} & $m_1$ & $m_2$ & $m_1$ & $m_2$ & $m_1$ & $m_2$ & $m_1$ & $m_2$ \\ \hline
			\multicolumn{1}{|c|}{30}  & 10 & 20 & 15 & 15 & 20 & 10 & 24 & 6 \\ 
			\multicolumn{1}{|c|}{40}  & 14 & 26 & 20 & 20 & 26 & 14 & 32 & 8 \\ 
			\multicolumn{1}{|c|}{50}  & 18 & 32 & 25 & 25 & 32 & 18 & 40 & 10 \\ \hline
		\end{tabular}
		\caption{\small Values of the sample sizes $m$, $m_1$ and $m_2$, according to each fixed percentage.}\label{ModelBasedTableSizes}
	\end{table}
	
	Table \ref{ModelBasedTableSummary} displays some of the frequentist properties of the estimators obtained by fitting the proposed disaggregated and aggregated models. In general, increasing the sample size leads us to smaller errors and variances (RW), so it is expected that errors associated with the disaggregated model's fit assuming $m = 50$ are smaller than the ones that consider $m = 30$ and $m = 40$. Note that, as the percentage of networks sampled in the first sampling stage decreases, the disaggregated model performs better, according to the RRMSEs e RAEs, since its values become smaller. Moreover, these error values associated with the aggregated model are higher than the ones obtained under the proposed model fit, except when $m=40$ and the sampling proportion is fixed in 80\%. The relative bias of the fitted MAs is smaller than the ones produced by the MDs models and both seem to underestimate the population totals $T$. The relative width of the proposed model for all scenarios are smaller than the ones provided by the aggregated model, even producing higher coverages. Overall, the disaggregated model presents a better performance than the aggregated model.
	
	\begin{table}[ht!]
		\centering
		{\small
			\begin{tabular}{|c|l|ccccc|}
				\cline{1-1} \cline{3-7}
				$m$                 &             & RRMSE & RAE & RB & RW & Cov. \\ \hline
				\multirow{5}{*}{30} & MD$_{35\%}$ & 0.334 & 0.263 & -0.030 & 1.027 & 100.00\\
				& MD$_{50\%}$ & 0.335 & 0.267 & -0.052 & 1.010 & 100.00\\
				& MD$_{65\%}$ & 0.343 & 0.278 & -0.082 & 1.030 & 100.00\\
				& MD$_{80\%}$ & 0.350 & 0.287 & -0.106 & 1.029 & 100.00\\
				& MA          & 0.357 & 0.283 & -0.015 & 1.119 & 99.00\\\hline
				\multirow{5}{*}{40} & MD$_{35\%}$ & 0.262 & 0.207 & -0.022 & 0.761 & 99.80\\
				& MD$_{50\%}$ & 0.267 & 0.214 & -0.042 & 0.762 & 99.60\\
				& MD$_{65\%}$ & 0.275 & 0.224 & -0.070 & 0.768 & 99.20\\
				& MD$_{80\%}$ & 0.302 & 0.252 & -0.110 & 0.791 & 98.40\\
				& MA          & 0.294 & 0.235 & -0.018 & 0.867 & 96.00\\\hline
				\multirow{5}{*}{50} & MD$_{35\%}$ & 0.226 & 0.178 & -0.015 & 0.619 & 99.80\\
				& MD$_{50\%}$ & 0.227 & 0.180 & -0.022 & 0.603 & 98.60\\
				& MD$_{65\%}$ & 0.235 & 0.190 & -0.045 & 0.614 & 98.60\\
				& MD$_{80\%}$ & 0.245 & 0.202 & -0.073 & 0.617 & 98.40\\
				& MA          & 0.252 & 0.204 & -0.015 & 0.697 & 90.80\\\hline
		\end{tabular}}\vspace{-0.2cm}
		\caption{\small Summary measurements of the point and 95\% credibility interval estimates for $T$ over 500 simulations obtained for the fits of the disaggregated and aggregated models, considering different sample sizes and proportions.} \label{ModelBasedTableSummary}\vspace{-0.3cm}
	\end{table}
	
	\begin{figure}[ht!]
		\begin{center}
			\includegraphics[width=15.5cm]{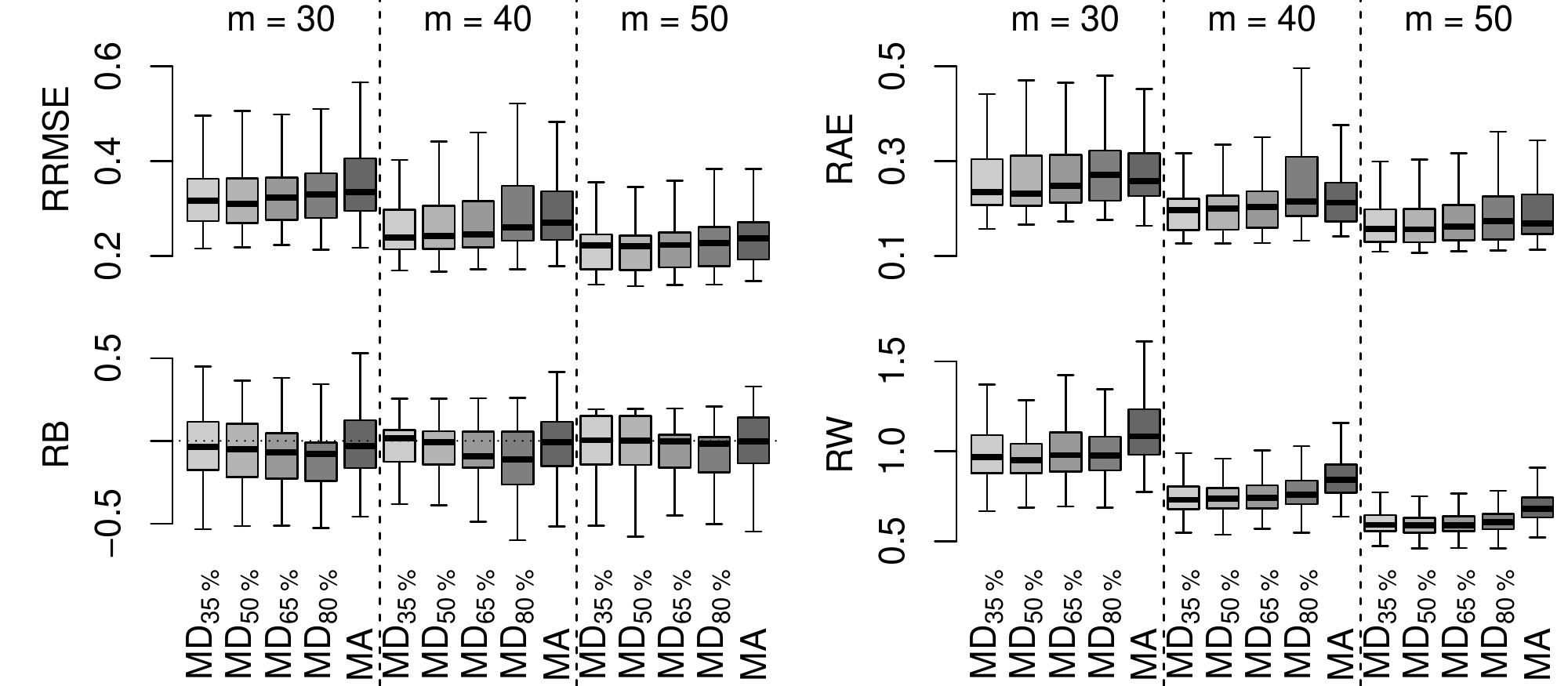}\vspace{-0.2cm}
			\caption{\small Boxplots with measurements of the point and 95\% credibility interval estimates for $T$ over 500 simulations obtained for the fits of the disaggregated and aggregated models, considering different sample sizes and proportions.}\label{ModelBasedBoxplot}
		\end{center}\vspace{-0.5cm}
	\end{figure}
	
	Figure \ref{ModelBasedBoxplot} presents the boxplot of some of the previous measurements associated with the estimates for $T$, and again it suggests that the disaggregated model performs better taking into account the variation of these values. In particular, there is an increase in RRMSEs and RAEs quartiles as we increase the percentage of networks sampled in the first sampling stage.
	
	\begin{figure}[ht!]
		\begin{center}
			\includegraphics[width=15.5cm]{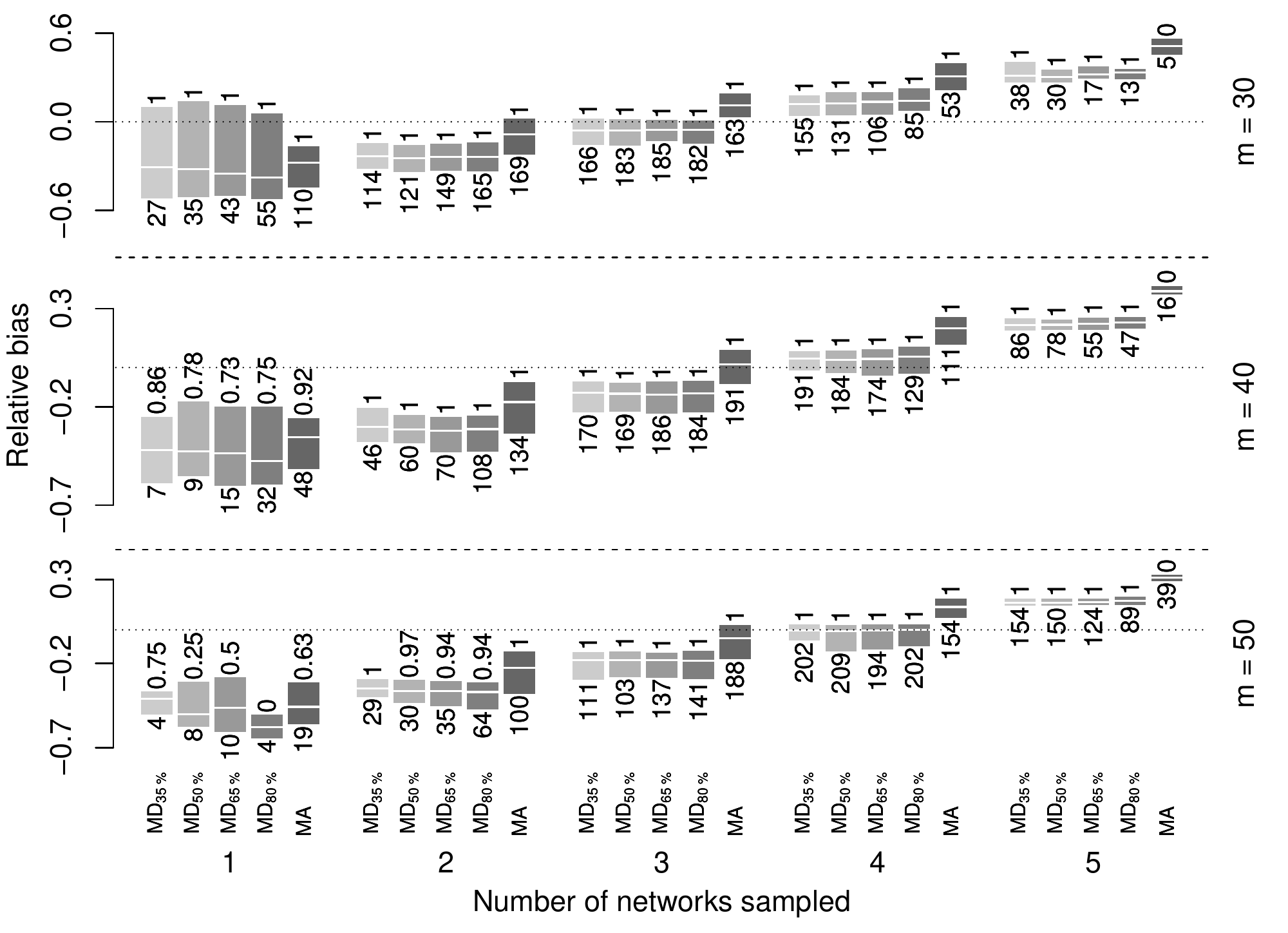}
			\caption{\small Mean (white line) and 2,5\% and 97,5\% quantiles (gray bars) of the relative bias estimates of $T$ over 500 simulations obtained for the fits of the disaggregated and aggregated models, considering different sample sizes, proportions, numbers of networks sampled (number of simulations given below the gray bars) and coverage (above the gray bars).}\label{ModelBasedBias}
		\end{center}\vspace{-0.8cm}
	\end{figure}
	
	To evaluate the effect of the number of networks sampled in the total estimates, Figure \ref{ModelBasedBias} shows some measurements of the relative bias estimates of $T$ from 500 simulations, according to each scenario and sampling methodology. For each scenario and sample size combination, the number of simulations distributed among the five possible quantities of sampled networks is given below the gray bars, with few simulations in some cases, and the coverage of the 95\% credibility intervals is shown above the gray bars. Remark that our proposal distribution may lead to none out-of-sample nonempty cells and, consequently, none out-of-sample networks, differently from \cite{Rapley2008}'s approach. Therefore, when we sample five networks, the aggregated model tends to overestimate more than the disaggregated models, and its coverage is zero regardless of the sample size $m$. Notice that for $m = 30$, all 95\% credibility intervals associated with disaggregated models contemplate the true value of $T$, while for $m = 40$, it happens when we sample more than one network and, for $m = 50$, when we sampled more than two networks. Moreover, among the disaggregated models, the model that provides the most coverage concerning all sample sizes is MD$_{35 \%}$.

	\begin{figure}[ht!]
		\begin{center}
			\includegraphics[width=15.5cm]{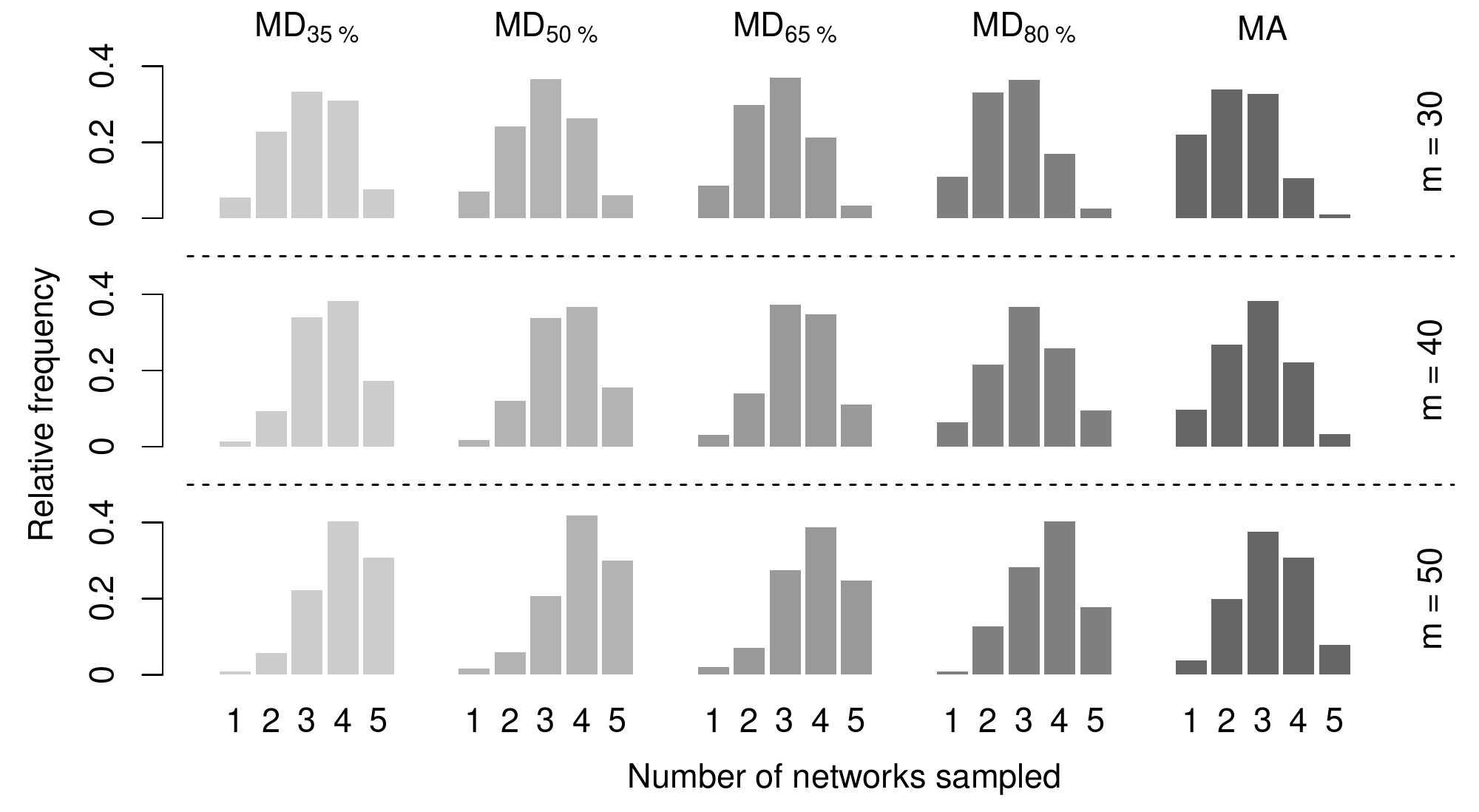}\vspace{-0.3cm}
			\caption{\small Relative frequency of the number of networks sampled over 500 simulations obtained for the fits of the disaggregated and aggregated models, considering different sample sizes and proportions.}\label{ModelBasedRedes}
		\end{center}\vspace{-1cm}
	\end{figure}
	
	Figure \ref{ModelBasedRedes} presents the barplots with the relative frequency of the number of networks sampled from 500 simulations, according to each scenario and sampling methodology. Note that, the proposed methodology provides a higher number of samples containing the entire population than the aggregated model and, in particular, as we decrease the percentage, the greater is the number of networks sampled. Moreover, as expected, we observed more nonempty sampled networks as we increase the sample size.
	
	Based on this study, the disaggregated model provides a more efficient sample, with a greater number of networks than the aggregated model. With relation to the estimators performance, the MD$_{35 \%}$ model showed to be more efficient and, therefore, we will concentrate on studying the properties of this model on an application to real data.
	
	\subsection{A real experiment}\label{DesignBasedExperiment}
	
	In this subsection, we analyze the performance of the disaggregated model with 35\% of $m$ networks sampled in the first sampling stage (MD$_{35\%}$) and the aggregated model (MA) using a real dataset. In order to simplify the notation we will refer MD$_{35\%}$ model as MD. The study variable considered is the number of African Buffaloes in an area of East Africa, while the auxiliary variable is the altitude (in meters). The choice of Buffalo and altitude was motivated by the fact that Buffaloes drink a lot of water (\cite{Prins1996}) and their spatial distribution depends on the prevailing climatic condition (\cite{Bennitt2014}), that is related to the altitude. Thus, areas with higher temperatures (lower altitude) lose terrain water (lakes or rivers) due to the evaporation, attracting little or no presence of Buffaloes.
	
	\begin{figure}[ht]
		\begin{center}
			\includegraphics[width=10cm]{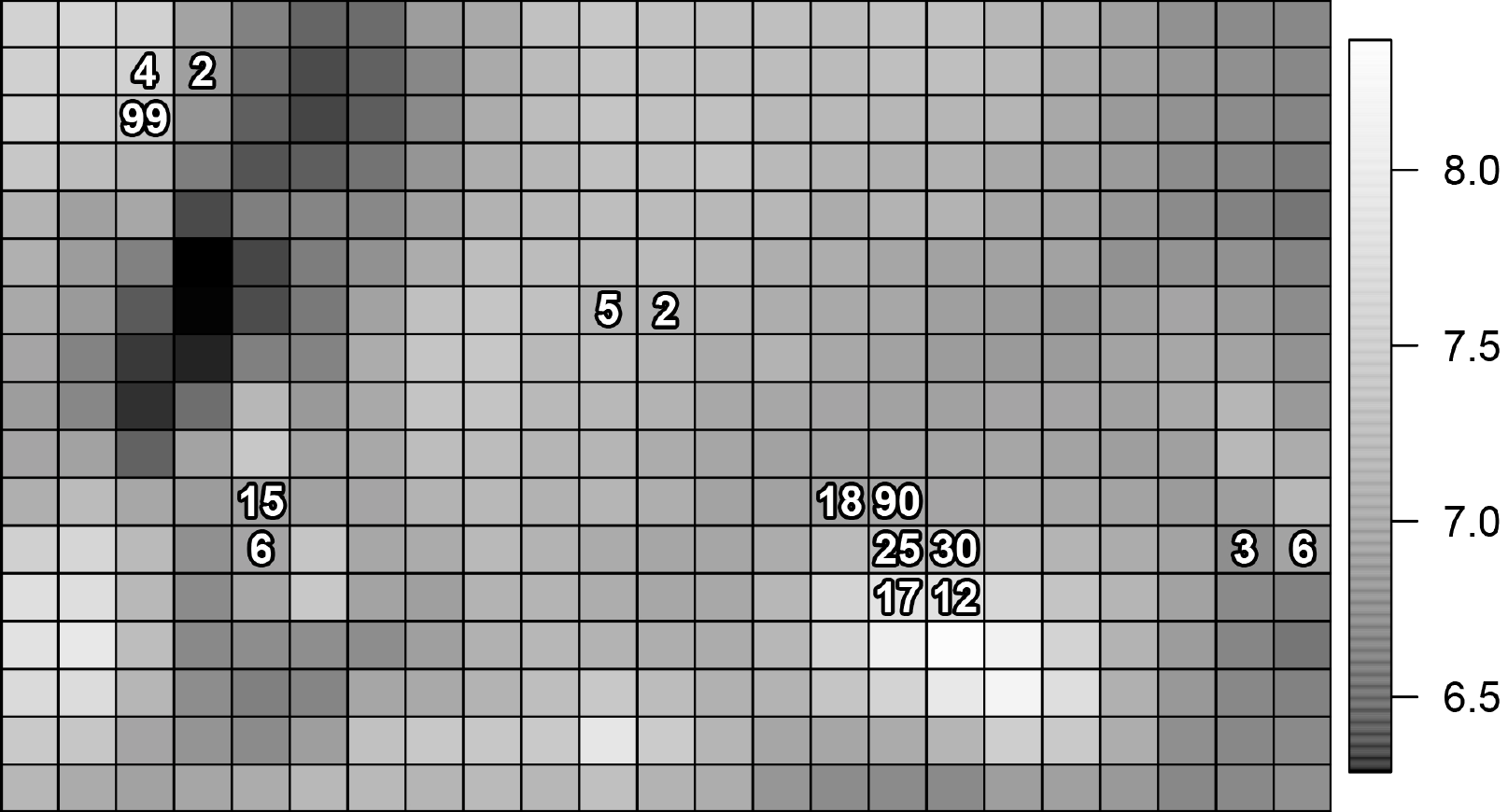}\vspace{-0.1cm}
			\caption{\small Altitude in a logarithm scale (gray background) and counts of African Buffaloes over parts of Kenya and Tanzania in 2010 in a grid with $M = 391$ cells.}\label{RealData}
		\end{center}\vspace{-0.6cm}
	\end{figure}
	
	The data on African Buffalo was obtained from maps produced from an aerial census. The census was conducted by the Kenya Wildlife Service, the Tanzania Wildlife Research Institute, and other partners during the wet season in the year 2010 covering an area of about 24,108$km^2$. The area covered was the Amboseli-West Kilimanjaro/Magadi-Natron cross border landscape, which covers parts of Kenya and Tanzania. The auxiliary data over the study area were obtained from the Shuttle Radar Topography Mission (STRM) database freely available for download from \url{https://www2.jpl.nasa.gov/srtm/}. In particular, we will use the altitude in a logarithm scale as covariate, since its values have a smaller order of magnitude. Figure \ref{RealData} presents the distribution and counts of the Buffalo in the study region along with pixels of auxiliary variables and shows that Buffaloes are mostly found in areas of higher altitudes.
	
	\begin{figure}[ht]
		\begin{center}
			\includegraphics[width=15.5cm]{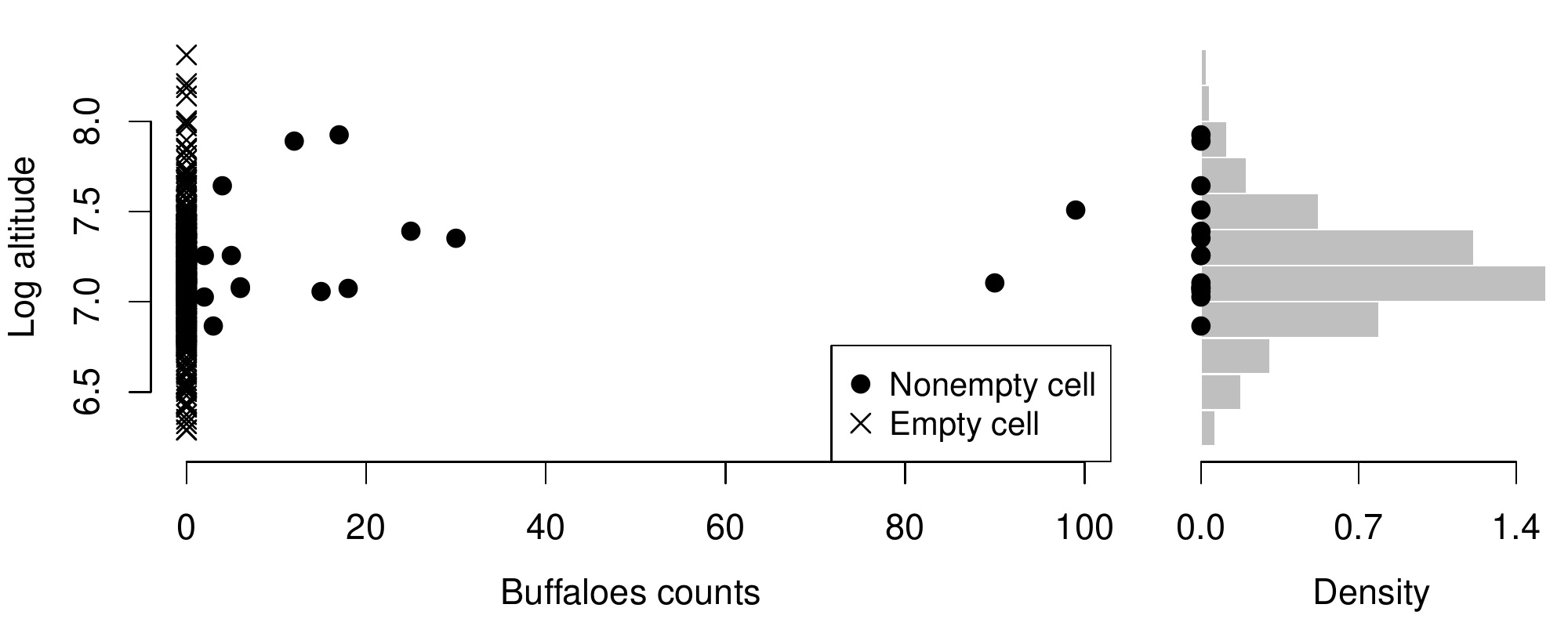}\vspace{-0.3cm}
			\caption{\small Plot of altitude in a logarithm scale versus counts of African Buffaloes, and covariate's histogram, where the black points in the histogram represent the nonempty cells.} \label{RealDataCovariate}
		\end{center}\vspace{-0.8cm}
	\end{figure}
	
	From Figure \ref{RealDataCovariate}, we can notice that Buffaloes tend not to be in areas in which the associated covariate has extreme values, i.e. they are concentrated in areas with intermediate log altitude values.
	
	Based on the relationship between the Buffalo counts and altitude, it seems natural to add the square of the covariate as another explanatory covariate, allowing us to model more accurately the effect of log altitude, which has a non-linear relationship with the Buffalo counts. Since we are using highly correlated covariates, centering them is helpful for the numerical schemes to converge. Thus, let $v(c)$ be the centered log altitude, for all cell $c \in R$, and $v^2(c)$ its respective square. Now, the covariate vector associated with the cell $c$ in the proposed model (\ref{Model}) is given by $\bfv_c$, for all $c \in R$. Introducing the squared log altitude leads us to consider the high correlation between covariates in the proposal distribution, which is detailed in Subsection \ref{FullConditionalThetas}.
	
	Remember that in the allocation process, described in Subsection \ref{ModelInference}, the available cells of $R$ are sampled with probability proportional to $\bflambda$, the set of intensities assigned to $R$ cells, which is obtained through information from sampled nonempty cells. Note that the real population used in this section (see Figure \ref{RealData}) is extremely rare, with small networks. Thus, the sample may contain a few number of nonempty cells and, consequently, few information to estimate $\bflambda$. Therefore, in this application, the weights used in the allocation process will be the probability of each cell not being empty, which is estimated from all sampled cells (empty and non-empty). Let $\phi(c) = 1$, if $c$ is a nonempty grid cell and $\phi(c) = 0$, otherwise. Define $\nu(c) = P(\phi(c) = 1)$, the probability that $c$ is a nonempty cell, and $\bfnu=(\nu(c_1), \dots, \nu(c_M))'$ the set of these probabilities, for all cell $c$ of $R$. Thus, the available cells of $R$ are sampled with probability proportional to $\bfnu$. To obtain these probabilities, the following structure will be included in the proposed model (\ref{Model}):\vspace{0.2cm}
	\begin{eqnarray}\label{ModelAplication}
	\phi(c)\mid \bfv_c,\bfrho&\sim&\mbox{Bernoulli}\left(\frac {1}{1+\exp\{-\bfv_c'\bfrho\}}\right), \qquad c \in R,
	\end{eqnarray}
	
	\noindent where $\bfrho = (\rho_0, \rho_1, \rho_2)'$ represents the regression coefficients vector associated with the covariates $\bfv_c = (1, v(c), v^2(c))'$. Note that, to estimate $\bfrho$, we will use the information from all empty and nonempty sample cells, unlike the $\bftheta$ adjust, which only takes into account nonempty cells of the sample. The full conditional posterior distribution of $\bfrho$ and the methods adopted to sample from it are detailed in Appendix \ref{AppendixA}, Subsection \ref{FullConditionalRhos}.
	
	Following the Bayesian paradigm, it is also assumed independent priors for the unknown parameters $\bfrho$ and its marginal prior distributions is denoted by $[\bfrho]$. Let $[\bfrho]$ be a non-informative prior with a zero-mean vector and covariance matrix $\sigma^2_\rho\bfI_{k+1}$, where $\bfI_{k+1}$ denotes the $(k+1)$-dimensional identity matrix. Since the African Buffaloes population is extremely sparse and clustered, we will set the prior distribution parameters of $\alpha$, presented in Subsection \ref{ModelInference}, as $a_\alpha = 3$ and $b_\alpha = 50$, to reflect the fact that $\alpha$ is necessarily small in this population. For $\beta$, we maintain the prior parameters set as $a_\beta = 1$ and $b_\beta = 9$.
	
	The study consists of drawing 500 samples of $m = 40$ networks of the real population according to each method. The proposed sampling methodology is divided into two stages, where the sample consists of 35\% of $m=40$ networks sampled in the first stage, this is $m_1 = 14$ networks are sampled randomly and $m_2 = 26$ networks based on weights $\bfomega$. In this study, we are omitting samples that consist of only empty networks since our proposed model requires at least one nonempty network in the first sample to adjust the weights $\bfomega$ properly.
	
	\begin{figure}[ht!]
		\begin{center}
			\includegraphics[width=15.5cm]{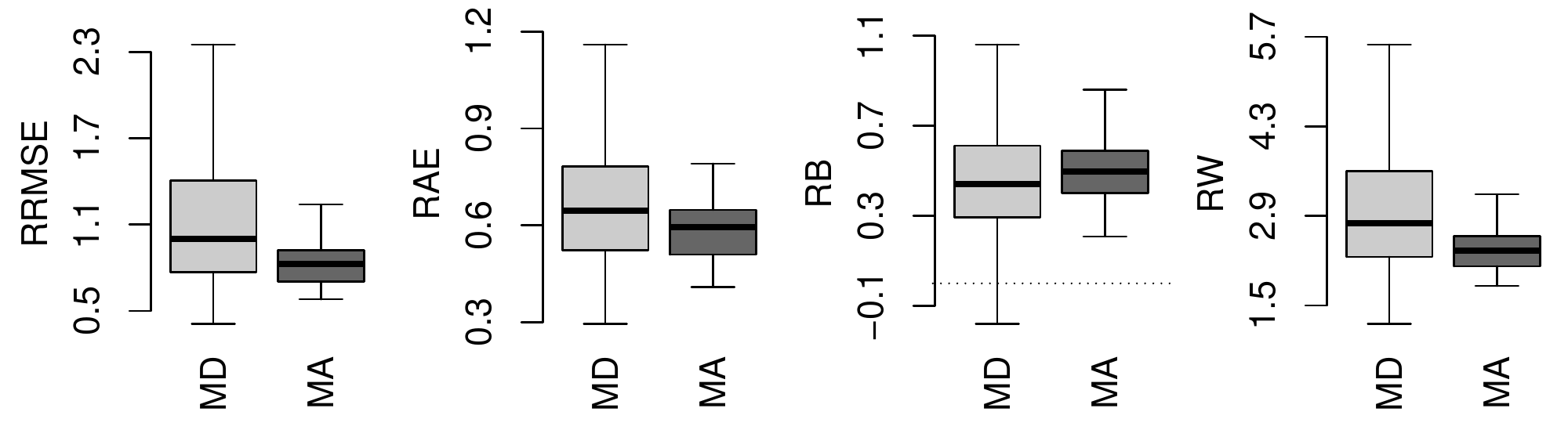}\vspace{-0.3cm}
			\caption{\small Boxplots with measurements of the point and 95\% credibility interval estimates for $T$ over 500 simulations obtained for the fits of the disaggregated and aggregated models to real data.}\label{DesignBasedBoxplots}
		\end{center}\vspace{-0.9cm}
	\end{figure}
	
	Figure \ref{DesignBasedBoxplots} presents the boxplots of the RRMSE, RAE, RB and RW of the Bayes estimators obtained from the disaggregated and aggregated models' fits, based on 500 simulations. Note that these measures' distributions under the disaggregated model are wider and higher than under the aggregated model.
	
	On the other hand, since the African Buffaloes population is extremely sparse and clustered, many samples consist either of networks with few cells each. These samples are expected to be of limited use in accurately estimating the population total. Thus, the results must ultimately be affected. Table \ref{DesignBasedTableNumber} shows that 79.4\% of the simulations based on MD and 84.4\% based on MA contain one or two nonempty networks sampled. Moreover, the aggregated model did not sample five nonempty networks once. Thus, in order to make the comparison more realistic it would be interesting to fix the number of nonempty networks sampled when each method is used.
	\vspace{0.3cm}
	
	\begin{table}[ht!]
		\centering
		\begin{tabular}{|c|ccccc|}
			\hline
			Percentage of nonempty networks sampled & 1 & 2 & 3 & 4 & 5 \\ \hline
			Disaggregated model & 35.4 & 44.0 & 18.0 & 2.2 & 0.4 \\
			Aggregated model & 36.4 & 48.0 & 13.6 & 2.0 & 0.0 \\ \hline
		\end{tabular}\caption{\small Percentage of networks sampled over 500 simulations under each model.}
		\label{DesignBasedTableNumber}
	\end{table}
	
	By this way, to facilitate the comparison between the models' fit, we repeated the study with 100 samples of the real population according to each method, fixing the final number of nonempty networks sampled, that is, we will consider 100 simulations with one to five nonempty networks sampled. Henceforth, we will refer to the number of nonempty networks sampled simply as number of networks sampled.
	
	Table \ref{DesignBasedTableFinal} displays some of the frequentist properties of the estimators for $T$ obtained by fitting the disaggregated and aggregated models, for each number of networks sampled. 
	When the number of networks sampled is fixed in four or five, the proposed model performs better than the aggregated one according to all the criteria. In general, the relative bias associated with the disaggregated model are smaller than the ones produced by the aggregated model, except when we have two networks sampled. Also, the coverages of the proposed model are higher than the aggregated model ones and, as we increase the number of networks sampled, the MA's coverage becomes smaller. In particular, with five networks sampled, none of the 500 95\% credibility intervals associated with the aggregated model contemplate the true value of $T$.\vspace{0.2cm}
	
	\begin{table}[ht!]
		\small
		\centering
		\begin{tabular}{c|cc|cc|cc|cc|cc|}
			\cline{2-11}
			& \multicolumn{10}{|c|}{Number of networks sampled}\\ \cline{2-11}
			& \multicolumn{2}{c|}{1} & \multicolumn{2}{c|}{2} & \multicolumn{2}{c|}{3} & \multicolumn{2}{c|}{4}&
			\multicolumn{2}{c|}{5}\\ \cline{2-11}
			& MD & MA & MD & MA & MD & MA & MD & MA & MD & MA\\ \hline
			\multicolumn{1}{|c|}{RRMSE} & 1.074 & 0.869 & 0.823 & 0.809 & 0.774 & 0.893 & 0.794 & 1.056 & 0.757 & 1.311\\
			\multicolumn{1}{|c|}{RAE} & 0.662 & 0.610 & 0.578 & 0.621 & 0.538 & 0.696 & 0.572 & 0.855 & 0.550 & 1.117\\
			\multicolumn{1}{|c|}{RB} & 0.339 & 0.538 & 0.374 & 0.364 & 0.428 & 0.585 & 0.549 & 0.832 & 0.550 & 1.117\\
			\multicolumn{1}{|c|}{RW} & 3.189 & 2.561 & 2.442 & 2.069 & 2.207 & 2.186 & 2.079 & 2.387 & 1.895 & 2.593\\
			\multicolumn{1}{|c|}{Cov.} & 100.00 & 100.00 & 100.00 & 64.00 & 100.00 & 51.00 & 100.00 & 32.00 & 96.00 & 0.00\\	\hline
		\end{tabular}
		\caption{\small Summary measurements of the point and 95\% credibility interval estimates for $T$ over 100 simulations for different numbers of networks sampled, obtained for the fits of the disaggregated and aggregated models.}\label{DesignBasedTableFinal}\vspace{-0.2cm}
	\end{table}
	
\begin{figure}[ht!]
		\begin{center}
			\includegraphics[width=15.5cm]{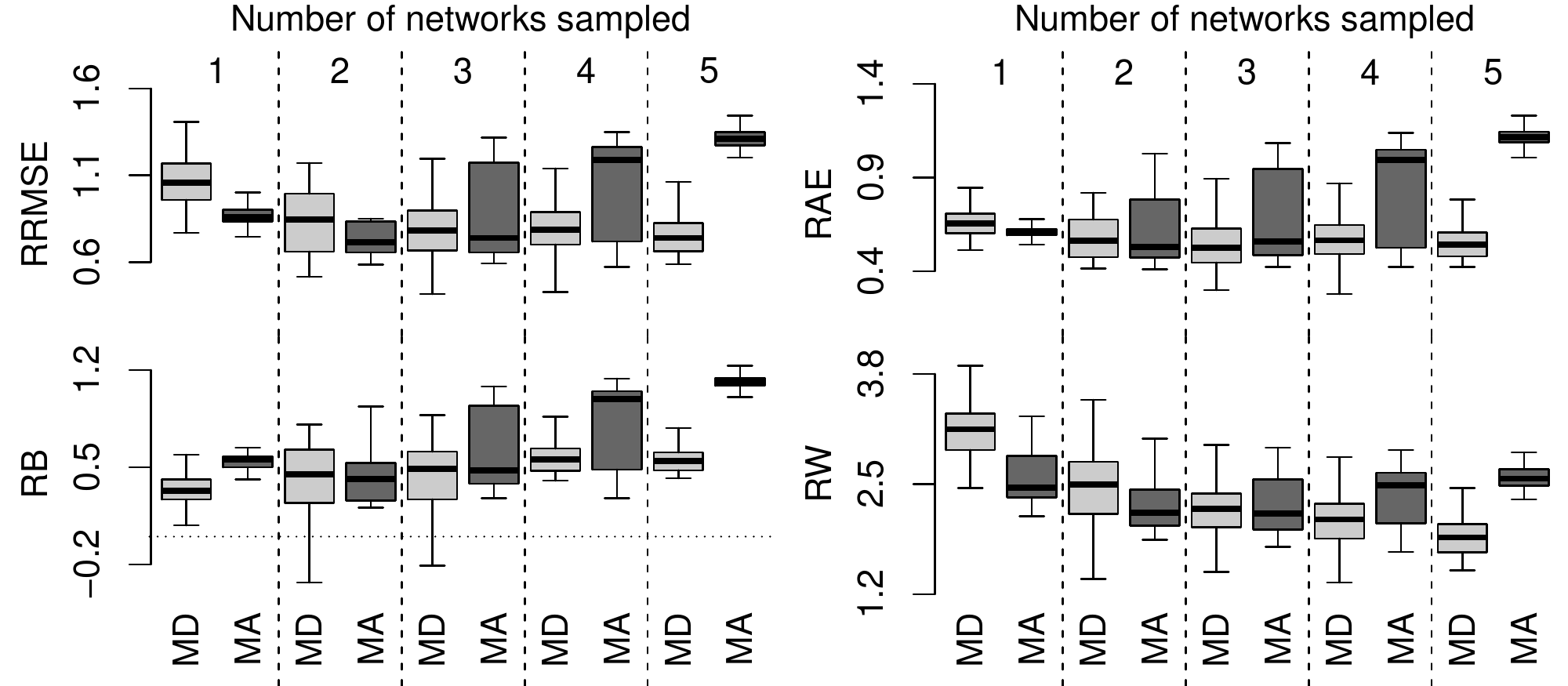}
			\caption{\small Boxplots with measurements of the point and 95\% credibility interval estimates for $T$ over 100 simulations for each number of networks sampled, obtained for the fits of the disaggregated and aggregated models to real data.}\label{DesignBasedBoxplotsFinal}
		\end{center}\vspace{-0.5cm}
	\end{figure}

	Figure \ref{DesignBasedBoxplotsFinal} presents the boxplots of some of the previous summary measurements for $T$, and the conclusion is analogous to the previous one. In particular, there is a decreasing behavior of the disaggregated model's relative widths as we increase the number of networks sampled.
	
	Finally, a summary comparison of population total $T$ estimators considering the disaggregated and aggregated models is presented, using RRMSE, RAE, RB and RW are presented in Table \ref{DesignBasedTableAggregated}, based on the 500 simulations resultant of the aggregation of 100 simulations with one to five networks sampled. Additionally, we compared these results to the ones obtained by applying an unbiased Raj’s estimator, detailed on \cite{Salehi1997}. This estimator of the population total is based only on the information contained in the selected networks, i.e., ignoring the information in the border cells. In this case, we used a normal approximation to set the 95\% confidence interval to the population total. Table \ref{DesignBasedTableAggregated} shows that both estimators have larger RRMSEs, RAEs and RBs than our proposed estimator, although it is well-known that Raj’s estimator is unbiased. Raj’s estimator has a much larger variance than its counterparts. The aggregated model produces 95\% credibility intervals that have lower nominal coverages than the others. Furthermore, our proposed model appears to be more efficient when applied to these data.
	
	\begin{table}[ht!]
		\centering 
		\begin{tabular}{c|ccccc|}
			\cline{2-6}
			\multicolumn{1}{c|}{}& RRMSE & RAE & RB & RW & Cov.\\ \hline 
			\multicolumn{1}{|c|}{Disaggregated model} & 0.845 & 0.580 & 0.448 & 2.362 & 99.2\\
			\multicolumn{1}{|c|}{Aggregated model} & 0.987 & 0.780 & 0.687 & 2.359 & 49.4\\
			\multicolumn{1}{|c|}{Raj's estimator} & 1.018 & 1.018 & 0.967 & 4.222 & 99.0\\ \hline
		\end{tabular}
		\caption{\small Summary measurements of the point and interval estimates of the total population,
			obtained by fitting the disaggregated and aggregated models and Raj’s estimator.}\label{DesignBasedTableAggregated}\vspace{-0.4cm} 
	\end{table}
	
\begin{figure}[ht!]
		\begin{center}
			\includegraphics[width=15.5cm]{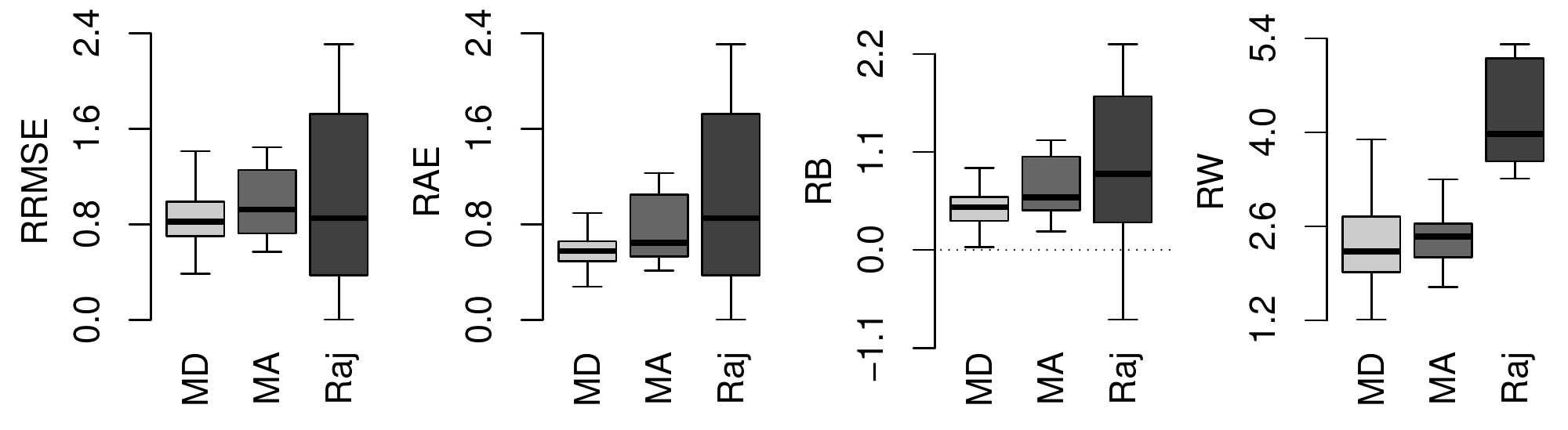}\vspace{-0.3cm}
			\caption{\small Boxplots with measurements of the point and 95\% credibility interval estimates for $T$ over 100 simulations for each number of networks sampled, obtained for the fits of the disaggregated and aggregated models.}\label{DesignBasedBoxplotsAggregated}
		\end{center}\vspace{-0.6cm}
	\end{figure}

	Figure \ref{DesignBasedBoxplotsAggregated} shows the boxplots of some measurements of the Bayes estimators obtained when fitting each model and Raj’s estimator. Here again, we see that the conclusion is analogous to the other.
	
	\begin{figure}[ht!]
		\centering
		\begin{tabular}{@{}c@{}} \subfigure[One network sampled]{
				\includegraphics[width=7cm]{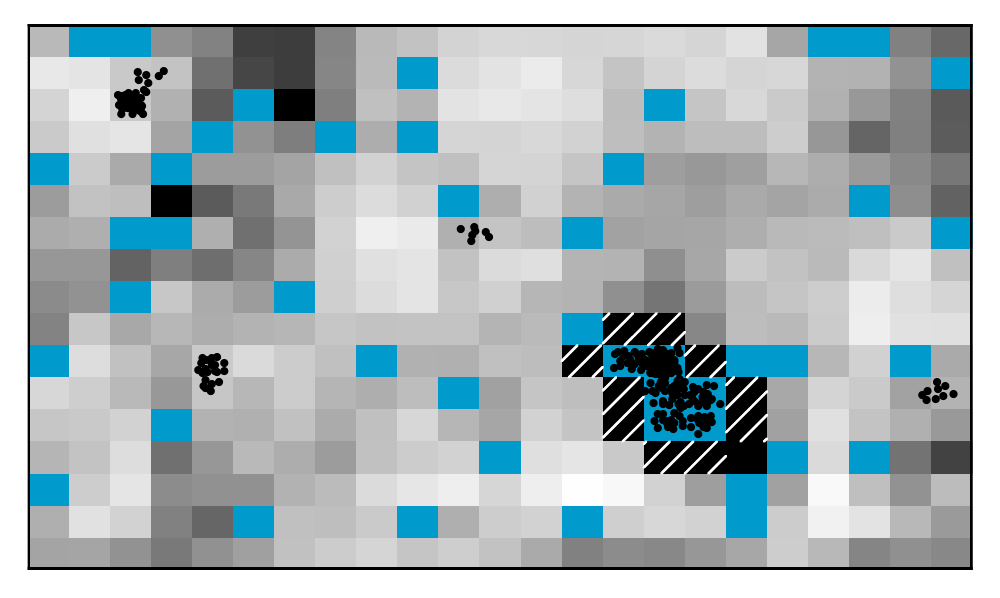}}\end{tabular}
		\quad
		\begin{tabular}{@{}c@{}} \subfigure[Two networks sampled]{
				\includegraphics[width=7cm]{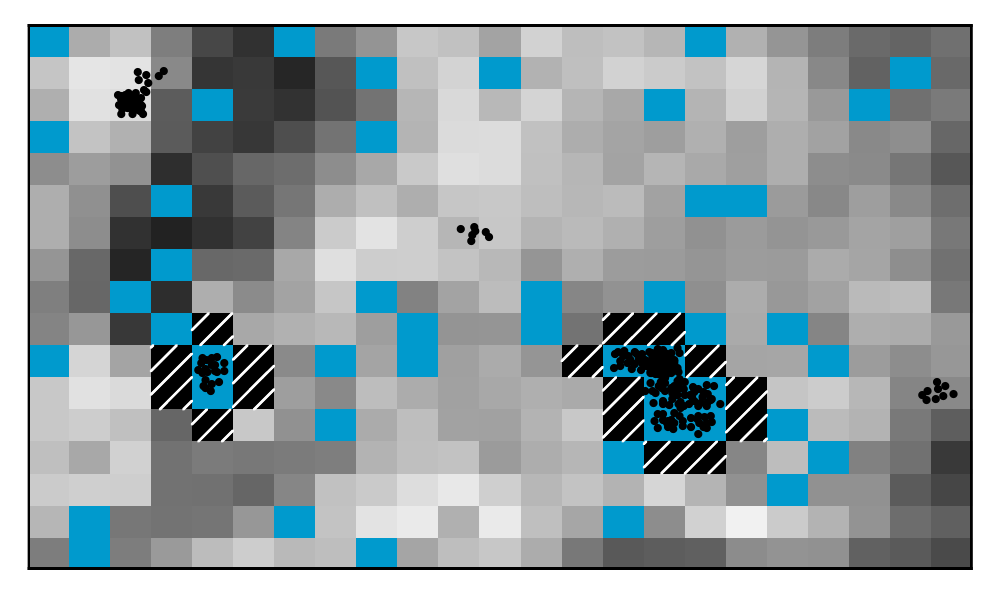}}\end{tabular}
		\begin{tabular}{@{}c@{}} \subfigure[Three networks sampled]{
				\includegraphics[width=7cm]{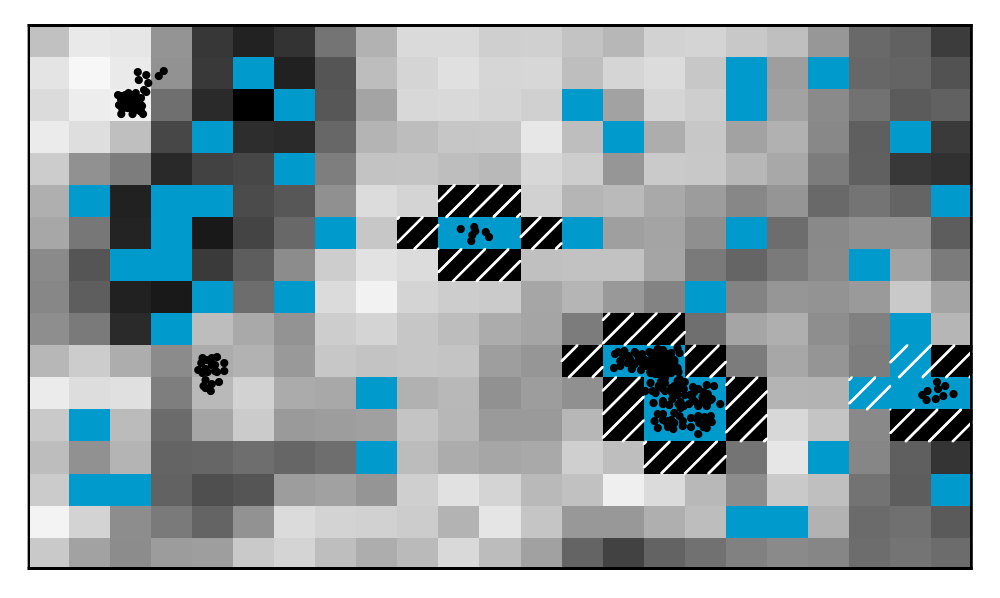}}\end{tabular}
		\quad
		\begin{tabular}{@{}c@{}} \subfigure[Four networks sampled]{
				\includegraphics[width=7cm]{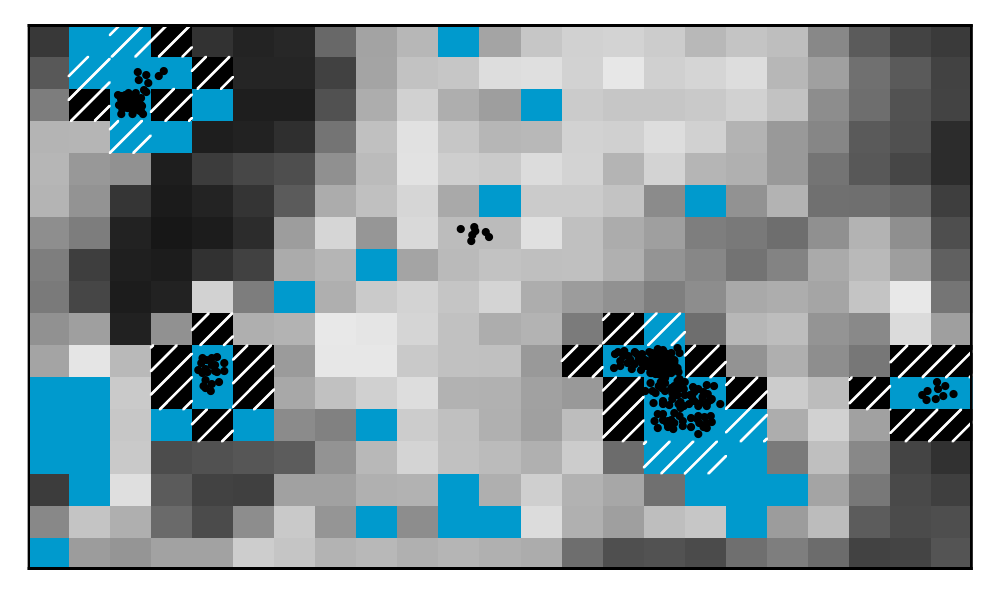}}\end{tabular}
		\begin{tabular}{@{}c@{}} \subfigure[Five networks sampled]{
				\includegraphics[width=7cm]{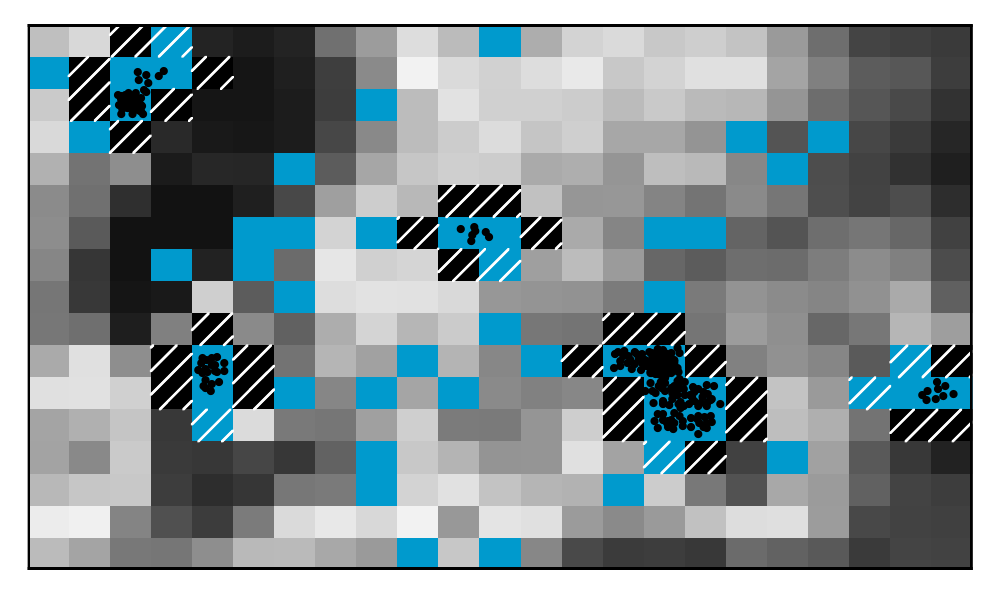}}\end{tabular}
		\vspace{-0.2cm}
		\caption{\small Maps of the posterior mean of African Buffalo counts $\eta(c)$ (gray background) for all out-of-sample cell $c$ of $R$, for each number of sampled networks, and its population (points) distributed in a region with 391 cells. The lighter the cells' color, the higher the posterior mean of that cell. The blue cells are the sampled networks' cells and the hatched cells correspond to the nonempty networks' border cells.}
		\label{DesignBasedMaps}\vspace{-0.2cm}
	\end{figure}
	
	In addition to other advantages previously seen, incorporating covariates into the model allows us to refer the out-of-sample nonempty cells spatially. Figure \ref{DesignBasedMaps} presents, for one sample of each number of networks sampled, a map of the posterior mean of the African Buffalo counts, i.e., the posterior mean of $\eta(c)$ for all out-of-sample cell $c$ of $R$. Remark that the lighter the cells' color, the higher the posterior mean of that cell. Due to the allocation process, the count estimates associated with the sampled nonempty networks' border cells are equal to zero (hatched black cells). Note that, the maps become darker as we increase the number of sampled networks, i.e. samples with more nonempty networks tend to estimate lower out-of-sample counts. Moreover, considering one to four networks sampled, the out-of-sample nonempty networks are located in lighter areas, indicating that the model can predict where they are. In the case with five networks sampled, due to the proposed model structure and the relation between the Buffalo counts and the covariate, we believe that the lighter areas would be more conducive to the establishment of new populations, although there are none out-of-sample nonempty networks.
	
	Figure \ref{Trace} in Appendix \ref{AppendixB} shows the trace plot with the posterior distribution of parameters $\alpha$ and $\beta$ and the population total $T$ when fitting the disaggregated model for one of the samples selected for each number of networks sampled. Table \ref{TableTrace} in Appendix \ref{AppendixB} presents the values of the Geweke criteria. Analyzing Figure \ref{Trace} and Table \ref{TableTrace} leads us to conclude that convergence appears to have been reached. The same conclusion was achieved for all 600 samples selected from this population.
	
	\section{Model-based experiment under different settings}\label{SimulationStudy}
	
	To examine the proposed methodology's performance under several scenarios, 500 populations were generated considering different configurations for each one of the four scenarios considered, which were created by varying the values of parameters $(\alpha, \beta)$. In particular, populations were simulated for 4 pairs of $(\alpha, \beta)$, with $\alpha, \beta \in \{0.10, 0.15\}$, which were set to create different degrees of rare and clustered populations. Then, an adaptive cluster sample of final size $m = 40$ was selected from each population with a 35\% proportion being sampled randomly, i.e. the first stage's sample size is $m_1 = 14$ and the second's one is $m_2 = 26$.
	
	Table \ref{SimulationStudyTable} shows summary statistics with some frequentist measurements of the posterior distributions of the model parameters for each of the four evaluated scenarios. Note that, the less rare and clustered the population is, the narrower the 95\% credibility interval is, and the greater the tendency for the model to underestimate its parameters. Moreover, the RRMSEs and RAEs do not vary much. In addition, the rarer and more clustered the population, the greater the coverage of the 95\% credibility intervals for the population total $T$, while the coverage for $\alpha$ and $\beta$ are close to the nominal level.
	
\begin{figure}[ht!]
		\begin{center}
			\includegraphics[width=15.5cm]{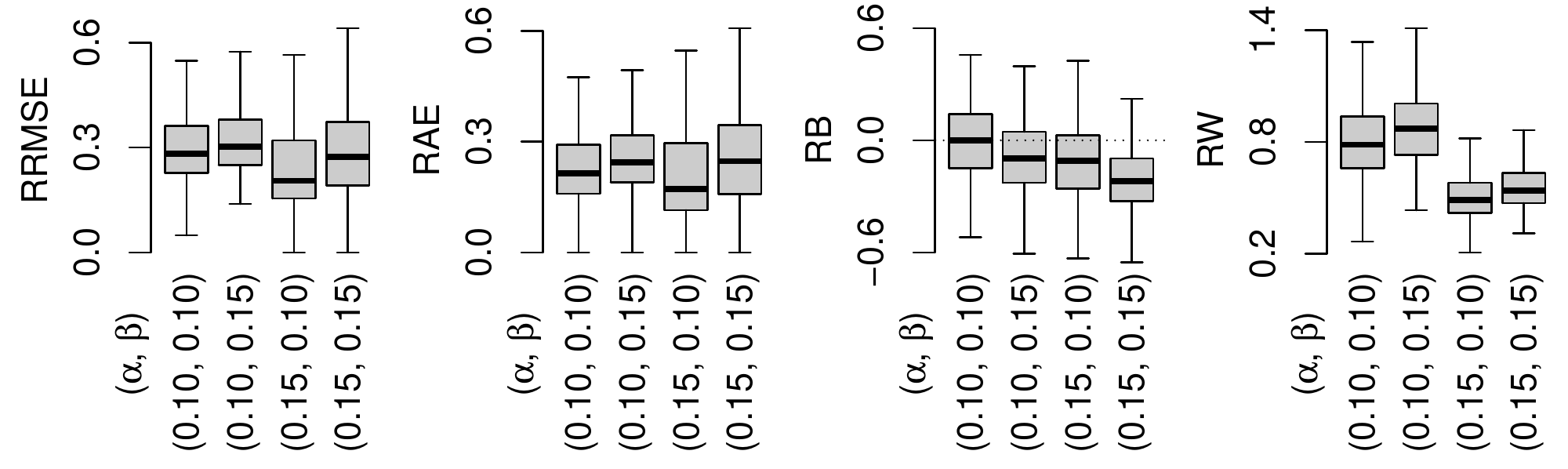}\vspace{-0.3cm}
			\caption{\small Boxplots with measurements of the point and 95\% credibility interval estimates of the proposed model and population parameters over 500 simulations for different values of $\alpha$, $\beta$ and $T$.}\label{SimulationStudyBoxplot}
		\end{center}\vspace{-0.7cm}
	\end{figure}

	Figure \ref{SimulationStudyBoxplot} presents the boxplots of some of the previous summary measurements for $T$. Note that, fixing $\alpha$, as we switch $\beta$ from 0.10 to 0.15, the RRMSEs, RAEs and RWs increase. Additionally, $T$ is slightly underestimated as the population becomes rarer and more clustered. 
	
	\begin{table}[ht!]
		\small
		\centering
		\begin{tabular}{c|ccc|ccc|}
			\cline{2-7}
			& $T$           & $\alpha$          & $\beta$          & $T$           & $\alpha$          & $\beta$ \\
			\cline{2-7}
			& \multicolumn{3}{c|}{$(\alpha, \beta) = (0.10, 0.10)$} & \multicolumn{3}{c|}{$(\alpha, \beta) = (0.10, 0.15)$} \\
			\hline
			\multicolumn{1}{|c|}{RRMSE} & 0.306 & 0.352 & 0.562 & 0.321 & 0.340 & 0.467\\
			\multicolumn{1}{|c|}{RAE} & 0.238 & 0.288 & 0.465 & 0.262 & 0.279 & 0.393 \\
			\multicolumn{1}{|c|}{RB} & 0.006 & 0.037 & -0.066 & -0.092 & -0.069 &  -0.146\\
			\multicolumn{1}{|c|}{RW} & 0.794 & 1.036 & 1.690 & 0.883 & 1.076 & 1.375\\
			\multicolumn{1}{|c|}{Cov.} & 95.00 & 93.60 & 95.00 & 99.00 & 98.80 & 95.00\\
			\hline
			& \multicolumn{3}{c|}{$(\alpha, \beta) = (0.15, 0.10)$} & \multicolumn{3}{c|}{$(\alpha, \beta) = (0.15, 0.15)$}\\ \hline
			\multicolumn{1}{|c|}{RRMSE} & 0.242 & 0.273 & 0.490 & 0.292 & 0.314 & 0.412\\
			\multicolumn{1}{|c|}{RAE} & 0.209 & 0.236 & 0.402 & 0.264 & 0.278 & 0.345\\
			\multicolumn{1}{|c|}{RB} & -0.114 & -0.109 & -0.022 & -0.216 & -0.209 & -0.126\\
			\multicolumn{1}{|c|}{RW} & 0.495 & 0.701 & 1.513 & 0.562 & 0.723 & 1.225\\
			\multicolumn{1}{|c|}{Cov.} & 89.60 & 93.80 & 97.00 & 86.40 & 91.60 & 96.00\\
			\hline
		\end{tabular}
		\caption{\small Summary measurements of the point and 95\% credibility interval estimates of the proposed model and population parameters over 500 simulations for different values of $\alpha$, $\beta$ and $T$.}\label{SimulationStudyTable}\vspace{-0.3cm}
	\end{table}
	
	Finally, considering the total population $T$, the scenario generated considering $\alpha = 0.15$ and $\beta = 0.10$, provided lower values of RRMSE and RAE errors, besides presenting a smaller relative width.
	
	\section{Conclusions and suggestions for future work}\label{Conclusion}
	
	We have considered the problem of estimating the total number of individuals in a rare and clustered population. A regular grid is superimposed on the interest region, placing the clusters, giving them a spatial size and allowing modeling the number of individuals selected by adaptive cluster sampling, as described by \cite{Salehi1997}, within this grid structure.
	
	Our approach is to model the observed counts of the selected grid cells and to use a model-based analysis to estimate the total population using the auxiliary information of covariates. To include this extra knowledge to the model, we proposed a model more flexible than the one introduced by \cite{Rapley2008}, since it models at cell level instead of network-level and assumes that the intensity in each cell of a cluster is related to the available covariates' values. Despite the higher computational cost, the proposed methodology considering a grid with 400 cells still runs on a home computer (CORE i7, 16GB) at an acceptable time (about 30 minutes on average).
	
	As evidenced by the simulated studies in Sections \ref{IntroductorySimulationStudy} and \ref{ComparisonAggregatedModel}, the incorporation of covariates into the model provided an improvement in the sampling process by getting more sampled networks. In practical situations, increasing the number of sampled networks directs and enhances the use of human and material resources, reducing expenses involved with the sampling procedure. Besides, it is also possible to spatially refer observed and unobserved networks, highlighting areas more conducive to the establishment of the studied population. Moreover, despite the challenges inherent in the spatial prediction problem - more specifically in the allocation of the networks and their counts through the interest region - the resulting maps showed to adequately indicate where the population under study is placed. We also performed changes in MCMC obtaining advances in inference: our proposal distribution may lead to none out-of-sample networks, and the performance of our Bayes estimator is substantially better than the one proposed by \cite{Rapley2008}.
	
	Simulation studies have assessed different scenarios varying the percentage of networks drawn in the first sampling stage and the parameters used to generate artificial populations. Our methodology has yielded satisfactory results and, in most cases, better than those obtained without additional information according to various comparison criteria. In the analyzed application, covariate information was successfully incorporated into the model by including quadratic terms in the linear predictor, evidencing the flexibility of the model to incorporate available auxiliary information. Real simulation studies have shown that the results are quite satisfactory according to several comparison criteria, validating the methodology proposed in this paper for practical situations. 
	
	The main findings of this work encourage an extension of this model to other spatial structures that reveal more information about the population.
	
	\spacingset{1}
	
	\spacingset{1.3}
	\appendix
	\section{Full conditional posterior distributions of the parameters in the proposed model}\label{AppendixA}
	
	In this section we present the posterior full conditional distributions of the components of the parameter vector $\bfTheta = (\bfeta_{\sbar},\bfY_{\sbar},P_{\sbar},X_{\sbar},\bftheta,\beta,\alpha)'$. We denote the posterior full conditional of a parameter $\phi$ in $\bfTheta$ by $[\phi\mid\cdots]$.
	
	\subsection{Full conditional posterior distribution of \texorpdfstring{$\bftheta$}{Lg}}\label{FullConditionalThetas}
	The posterior full conditional of $\bftheta = (\theta_0, \theta_1, \dots, \theta_k)'$ is proportional to
	$$
	[\bftheta\mid \cdots]\ \propto\
	\prod_{c \in \bfC_s}\frac{\exp\{-\exp\{\bfv_c'\bftheta\}+\eta(c)\bfv_c'\bftheta\}}{ 1-\exp\{-\exp\{\bfv_c'\bftheta\}\}}\times\exp\left\{-\frac{1}{2\sigma^2_\theta}\bftheta'\bftheta\right\},
	$$
	which does not have an analytical closed-form. We use the block Metropolis-Hastings algorithm with a multivariate Normal proposal, whose vector mean is the current value of the parameter and the covariance matrix is fixed at $\bfI_{k+1}\sigma^2_{\theta^*}$, where the term $\sigma^2_{\theta^*}$ controls the acceptance rates and $\bfI_{k+1}$ denotes the $(k+1)$-dimensional identity matrix.
	
	Considering the structure presented in Subsection \ref{DesignBasedExperiment}, where the squared log altitude is introduced to the model, the covariance matrix is fixed at $(\bfV'\bfV)^{-1}\sigma^2_{\theta^*}$, where the term $\sigma^2_{\theta^*}$ controls the acceptance rates and $\bfV$ is the matrix with rows $\bfv_c'$, for all nonempty cell $c$ of the sample, that is $c \in \bfC_s$.
	
	
	
	\subsection{Full conditional posterior distribution of \texorpdfstring{$\alpha$}{Lg}}\label{FullConditionalAlpha}
	The posterior full conditional of $\alpha$ is proportional to
	$$
	[\alpha \mid \cdots]\ \propto\ \frac{\alpha^{x_s+x_{\sbar}+a_\alpha-1}(1-\alpha)^{M-x_s-x_{\sbar}+b_\alpha-1}}{1-(1-\alpha)^M},
	$$
	which is close to a Beta distribution but is not truly a Beta distribution, due to the truncation term. We use the Metropolis-Hastings algorithm with a Beta proposal with parameters $\mbox{beta}(x_s+x_{\sbar}+a_\alpha,\ M-x_s-x_{\sbar}+b_\alpha)$.
	
	\subsection{Full conditional posterior distribution of \texorpdfstring{$\beta$}{Lg}}
	The posterior full conditional of $\beta$ is proportional to
	$$
	[\beta \mid \cdots]\ \propto\  \frac{\beta^{p_s+p_{\sbar}+a_\beta-1}(1-\beta)^{x_s+x_{\sbar}-p_s-p_{\sbar}+b_\beta-1}}{1-(1-\beta)^x},
	$$
	which is close to a Beta distribution but is not truly a Beta distribution, due to the truncation term. We use the Metropolis-Hastings algorithm with a Beta proposal with parameters $\mbox{beta}(p_s+p_{\sbar}+a_\beta,\ x_s+x_{\sbar}-p_s-p_{\sbar}+b_\beta)$.
	
	
	\subsection{Full conditional posterior distribution of \texorpdfstring{$(X_{\sbar},P_{\sbar},\bfY_{\sbar},\bfeta_{\sbar})$}{Lg}} \label{FullConditionalQuantities}
	The joint posterior full conditional of $(X_{\sbar},P_{\sbar},\bfY_{\sbar},\bfeta_{\sbar})$ is proportional to
	\begin{eqnarray*}
		&&[X_{\sbar},P_{\sbar},\bfY_{\sbar},\bfN_{\sbar},\bfeta_{\sbar}\mid \cdots]\\
		&\propto& \prod_{j=1}^{m} \sum_{c\in\bfG_{i_j,j}} \frac{\displaystyle  \pi(c)} {\displaystyle \sum_{c \in R}\pi(c) - \sum_{k=0}^{j-1}\sum_{c\in\bfe_{k}}\pi(c)}\times\prod_{c \in \bfC_{\sbar}}\frac{\exp\{-\exp\{\bfv_c'\bftheta\}+\eta(c)\bfv_c'\bftheta\}}{\eta(c)!( 1-\exp\{-\exp\{\bfv_c'\bftheta\}\}) } \\ 
		&&\times\left(\frac{1}{p_s+p_{\sbar}}\right)^{\sum_{i=1}^{p_s+p_{\sbar}}(y_i-1)}\frac{1}{(p_s+p_{\sbar})!}\frac{\beta^{p_{\sbar}}(1-\beta)^{x_{\sbar}-p_{\sbar}}}{1-(1-\beta)^{x_s+x_{\sbar}}} \frac{\alpha^{x_{\sbar}}(1-\alpha)^{x_{\sbar}}}{(M-x_s+x_{\sbar})!},
	\end{eqnarray*}
	which does not have an analytical closed-form. We use the Metropolis-Hastings algorithm for sampling $(X_{\sbar}, P_{\sbar}, \bfY_{\sbar}, \bfeta_{\sbar})$ jointly. From the proposal distribution, it is straightforward to sample $(X_{\sbar}, P_{\sbar}, \bfY_{\sbar}, \bfeta_{\sbar})$ and jointly accept or reject these values using (a) as the target distribution.
	
	It is useful to generate $X_{\sbar}$ from a discrete uniform distribution with support in the set $\{X^*\pm k: k = 1,\dots,5 \}$, where $X^*$ is is the current value of the $X$. Then, make $X_{\sbar} = X - X_s$, ensuring that $X_s < X < M$, since the number of nonempty cells in $R$ is at most $M$ and it's known that there are $X_s$ nonempty cells in $R$.  Note that, $P_{\sbar}$ is the number of non-empty networks formed out of the $X_{\sbar}$ non-empty grid cells. Then, $P_{\sbar}$ is generated by sampling from the truncated Binomial$(X_{\sbar}, \beta)$ distribution. Notice that $\bfY_{\sbar}$ is the number of non-empty grid cells in each of the $P_{\sbar}$ networks, so we generate $\bfY_{\sbar}$ from the $\bfone_{P_{\sbar}} + \mbox{multinomial}\left(X_{\sbar} - P_{\sbar},\frac{1}{P_{\sbar}}\bfone_{P_{\sbar}}\right)$ distribution. Then, the set of cells $\bfC_{\sbar}$ that compose the out-of-sample nonempty networks is established from $Y_{\sbar}$ allocation process, described in Subsection \ref{ModelInference}. From the covariates associated with $\bfC_{\sbar}$ cells, we generate $\bfeta_{\sbar}$ elements from the truncated Poisson$(\exp\{\bfv_c'\bftheta\})$ distribution, for $c \in \bfC_{\sbar}$. Therefore, the proposal distribution is
	\begin{eqnarray}
	[X_{\sbar},P_{\sbar},\bfY_{\sbar},\bfeta_{\sbar}]_{prop}&=&\frac{1}{10}
	\times\frac{x_{\sbar}!}{p_{\sbar}!}\frac{\beta^{p_{\sbar}}(1-\beta)^{x_{\sbar}-p_{\sbar}}}{1-(1-\beta)^{x_{\sbar}}}
	\times\prod_{i\notin s}\frac{1}{(y_i-1)!}\left(\frac{1}{p_{\sbar}}\right)^{y_i-1} \nonumber\\
	&&\times\prod_{c \in \bfC_{\sbar}}\frac{\exp\{-\exp\{\bfv_c'\bftheta\}+\eta(c)\bfv_c'\bftheta\}}{\eta(c)!( 1-\exp\{-\exp\{\bfv_c'\bftheta\}\}) }.\nonumber
	\end{eqnarray}
	
	The improved sampling procedure, detailed in Subsection \ref{SamplingProcedure}, leads us to draw a greater number of networks than without using the weights $\bfomega$. Therefore, the final sample $s$ (made up by the first and second samples) may include all networks from $R$. Thus, it is plausible that the proposal distribution of the disaggregated model fitted to the final sample may lead to none out-of-sample nonempty cells, that is $X_{\sbar} = 0$. Note that, in this case, the number of nonempty cells in $R$, $X$, can assume value $X_s$. So, in $X$ generation, make $X_{\sbar} = X - X_s$, but ensuring that $X_s \leq X < M$. If $X = X_{s}$, then $X_{\sbar} = 0$ and the quantities $P_{\sbar}$, $\bfY_{\sbar}$ and $\bfeta_{\sbar}$ are, necessarily, equal to zero.  Therefore, the proposal distribution when $X_{\sbar} = 0$ is
	$$
	[X_{\sbar},P_{\sbar},\bfY_{\sbar},\bfeta_{\sbar}]_{prop}\ =\ \frac{1}{10}.
	$$
	
	\subsection{Full conditional posterior distribution of \texorpdfstring{$\bfrho$}{Lg}}\label{FullConditionalRhos}
	The posterior full conditional of $\bfrho$ (from Subsection \ref{DesignBasedExperiment}) is proportional to
	$$
	[\bfrho\mid \cdots]\ \propto
	\prod_{c \in \{\bfC_s, \overline{\bfC}_s\}} \left(\frac {1}{1+\exp\{-\bfv_c'\bfrho\}}\right)^{\phi(c)}\left(\frac {\exp\{-\bfv_c'\bfrho\}}{1+\exp\{-\bfv_c'\bfrho\}}\right)^{1-\phi(c)}\times\exp\left\{-\frac{1}{2\sigma^2_\rho}\bfrho'\bfrho\right\},
	$$
	which does not have an analytical closed-form. We use the block Metropolis-Hastings algorithm with a multivariate Normal proposal, whose vector mean is the current value of the parameter and the covariance matrix is fixed at $(\bfV'\bfV)^{-1}\sigma^2_{\rho^*}$, where the term $\sigma^2_{\rho^*}$ controls the acceptance rates and $\bfV$ is the matrix with rows $\bfv_c'$, for all cells $c$ of the sample, that is $c \in \{\bfC_s, \overline{\bfC}_s\}$.
	
	
	\section{Assessment of MCMC with real data}\label{AppendixB}
	
\begin{figure}[ht!]
		\begin{center}
			\includegraphics[width=14.4cm]{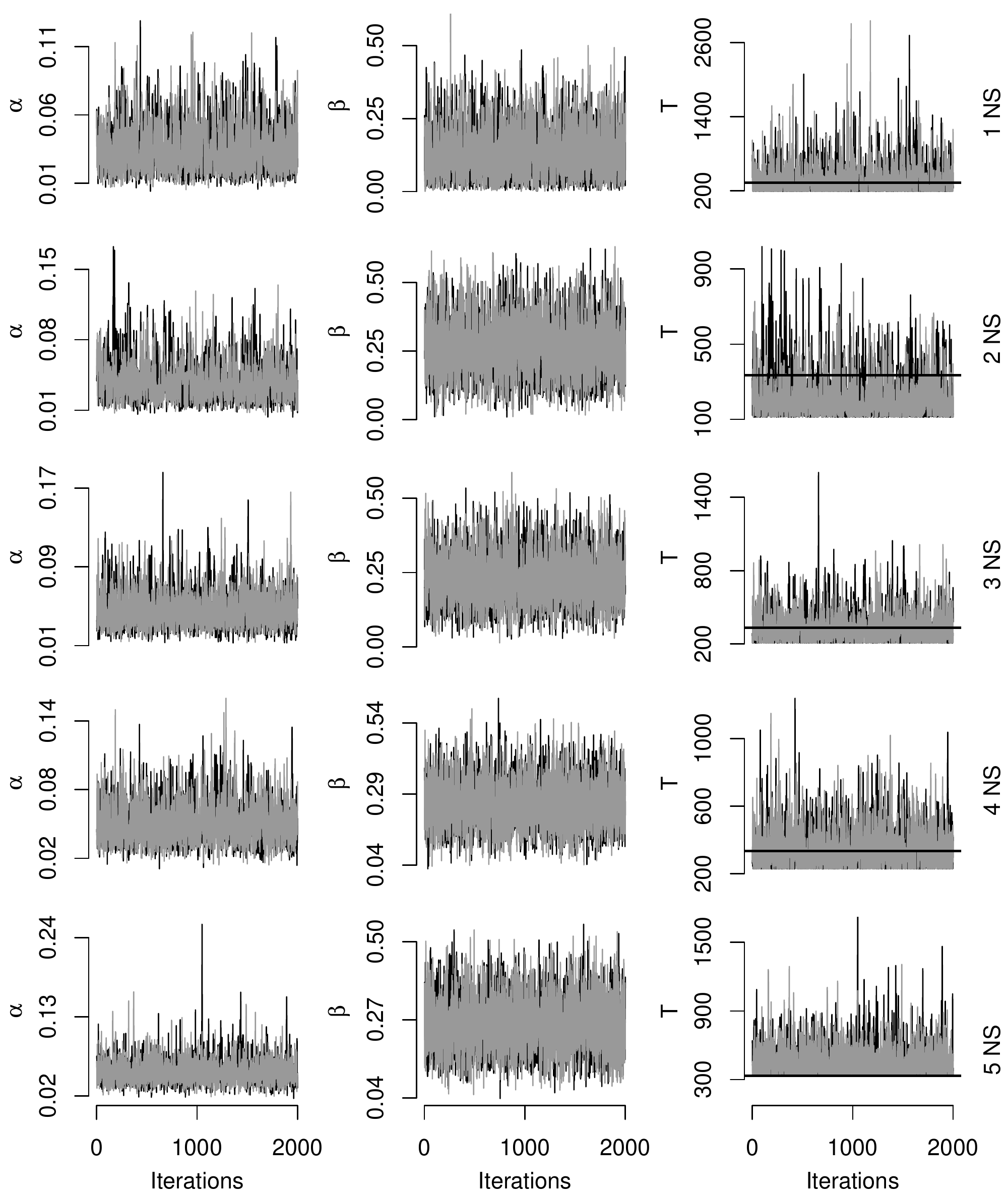}\vspace{-0.1cm}
			\caption{\small Trace plot with the posterior densities of $\alpha$, $\beta$ and $T$ obtained from the fits of the disaggregated and the aggregated models to real data, for each number of networks sampled (NS). The black line represents the true value of $T$.}\label{Trace}
		\end{center}
	\end{figure}

	In Section \ref{ComparisonAggregatedModel}, we compared the results of our approach to those obtained using the model proposed by \cite{Rapley2008}. This appendix presents the convergence results of the design-based experiment with a real population, displayed in Subsection \ref{DesignBasedExperiment}. We evaluated the convergence of two parallel chains according to each number of networks sampled from the real population. The results are presented in Table \ref{TableTrace} and Figure \ref{Trace}.
	
	\begin{table}[ht]
		\centering
		\begin{tabular}{c|ccccc|}
			\cline{2-6}
			& \multicolumn{5}{|c|}{Number of networks sampled} \\ \hline 
			\multicolumn{1}{|c|}{Parameter} & 1        & 2        & 3       & 4       & 5     \\ \hline
			\multicolumn{1}{|c|}{$\alpha$}  & -2.04 & 0.96 & -0.53 & -0.90 & 0.94\\
			\multicolumn{1}{|c|}{$\beta$}   & -2.18 & 1.50 & 1.14 & 0.20 & 0.54\\
			\multicolumn{1}{|c|}{$T$}       & -1.78 & 0.52 & -0.12 & -1.08 & 1.22\\ \hline
		\end{tabular}
		\caption{\small Geweke convergence diagnostic for some of the parameters estimated for the real population for both models.}\label{TableTrace}
	\end{table}
	
\end{document}